\documentclass[12pt,preprint]{aastex}

\usepackage{rotating}

\newcommand{\ie}{\mbox{i.e.}}

\newcommand{\NH}{\mbox{$N_{\rm H}$}}
\newcommand{\nh}{\mbox{$N_{\rm H~}$}}

\newcommand{\skipthis}[1]{}

\newcommand{\psqcm}{{\rm cm}^{-2}}

\newcommand{\ps}{{\rm s}^{-1}}

\newcommand{\erg}{{\rm ergs}}

\newcommand{\be}{\begin{equation}}
\newcommand{\ee}{\end{equation}}

\newcommand{\e}{et al.\ }

\def\ie{{i.e.\ }}

\slugcomment{\today}

\shorttitle{Young Suns of Orion}
\shortauthors{Wolk et al.}

\begin{document}


\title{Stellar Activity 
on the Young Suns of Orion:\\
COUP Observations of K5-7 Pre-Main Sequence Stars}


\author{S.J. Wolk 
and F.R. Harnden Jr.\altaffilmark{1} }
\affil{Harvard--Smithsonian Center for Astrophysics,
    60 Garden Street, Cambridge, MA 02138}
\email{swolk@cfa.harvard.edu
}
\author{E. Flaccomio \& G. Micela}
\affil{INAF - Osservatorio Astronomico G.S. Vaiana - Piazza Parlamento I, 90134 Palermo, Italy}

\author{F. Favata}
\affil{ Astrophysics Division -- Research and Space Science Support Department 
of ESA, ESTEC, Postbus 299, NL-2200 AG, Noordwijk, The Netherlands}


\author{H. Shang}
\affil{Institute of Astronomy and
Astrophysics, Academia Sinica, Taiwan}
\and

\author{E.D. Feigelson}
\affil{Department of Astronomy \& Astrophysics, Pennsylvania State
University, University Park, PA 16802}


\altaffiltext{1}{Universe Division, Science Mission Directorate NASA Headquarters}


\begin{abstract}
    In January 2003, the $Chandra$ Orion Ultradeep Project
(COUP) detected about 1400 young stars during a 13.2 day observation of
the Orion Nebula Cluster (ONC).  This paper is a study of the X-ray
properties of a well-defined sample of 28 solar-mass ONC stars based
on COUP data.  Our goals are to characterize the magnetic activity
of analogs of the young Sun and thereby to improve understanding of
the effects of solar X-rays on the solar nebula during the era of
planet formation.  Given the length of the COUP observation we are
able to clearly distinguish characteristic and flare periods for all
stars.
     We find that active young Suns spend 70\% of their time in a
characteristic state with relatively constant flux and magnetically
confined plasma with temperatures $ kT_2 \simeq 2.1 \times  kT_1$.
During characteristic periods, the $0.5-8$ keV X-ray luminosity is
about 0.03\% of the bolometric luminosity. One or two powerful
flares per week   with peak luminosities $\log L_x \sim 30-32$ ergs
s$^{-1}$ are typically superposed on this characteristic
emission accompanied by heating of the hot plasma component
from $\simeq 2.4$ keV to $\simeq 7$ keV at the flare peak. The energy
distribution of flares superposed on the
characteristic emission level follows the relationship $dN/dE \propto E^{-1.7}$.
The flare rates are consistent with the
production of sufficiently energetic protons to spawn a
spallogenic origin of some important short-lived radionuclides found
in ancient meteorites.  The X-rays can ionize gas in the
circumstellar disk at a rate of $6 \times 10^{-9}$ ionizations per
second at 1 AU from the central star, orders of magnitude
  above cosmic ray ionization rates.  The estimated energetic particle
  fluences are sufficient to
 account for many isotopic anomalies observed in meteoritic inclusions.

\end{abstract}




\section{Introduction}
The formation of stars and their planetary systems is
generally viewed as a low temperature phenomenon. Dark
molecular cores at $T \sim 10$ K collapse to pre-main
sequence (PMS) stars with surfaces at $T \sim 10^3$ K
surrounded by dusty disks with characteristic temperatures
of $T \sim 10^2$ K.  These earliest stages of stellar
evolution are studied primarily in the millimeter through
optical bands. Most efforts to understand the astrophysics
of these stages treat gravitational, hydrodynamical and
chemical processes of neutral molecular material in gaseous
and solid forms.

However, three lines of evidence point to the presence of
higher energy phenomena in young stellar systems,
ranging from excited gas at temperatures of $T \sim 10^4$ K, to
highly-ionized plasma at $T \sim 10^7$ K, to particles
accelerated up to $kT \sim$ MeV energies:
\begin{enumerate}

\item Low-excitation emission line gas, which dominates the
optical spectra of T Tauri stars and was formerly attributed to
hot outflowing winds, is now commonly thought to arise in
magnetically funneled accretion from the circumstellar
disk (see review by Hartmann 2001). To conserve
angular momentum, the accretion is accompanied by ejection
of disk material in magnetically collimated jets seen as
Herbig-Haro outflows (Shu \e 2000).

\item X-ray emission is ubiquitous in PMS stars at
levels that are orders of magnitude above those seen in most main
sequence stars.  These PMS stars generally show high-amplitude rapid
variability and the type of hard spectra that are associated with violent
magnetic reconnection flares when seen on the surface of the
Sun, dMe flare stars and other magnetically active
late-type stars (Feigelson \& Montmerle 1999, Favata \& Micela 2003, 
G{\" u}del 2004).

\item Studies of the isotopic composition of components of
ancient meteorites require either the injection of
short-lived nuclides into the molecular cloud that formed
our solar system, or {\it in situ} irradiation of solids by
MeV baryons in the solar nebula (Goswami \& Vanhala 2000). Radio
observations show gyrosynchrotron radiation
from MeV electrons accelerated by magnetic flares in some
young stellar systems (G{\" u}del 2002).
\end{enumerate}

Together, these studies indicate that astrophysical
modeling of young stellar systems, and particularly their
protoplanetary disks, requires treatment of ionization,
nuclear spallation, and magnetohydrodynamical processes in
addition to processes affecting neutral materials
(see reviews by Feigelson \& Montmerle 1999, 
Glassgold, Feigelson, \& Montmerle 2000; hereafter GFM00,
Feigelson 2005 and Glassgold \e 2005). 
High-energy photon and particle irradiation of
protoplanetary disks may substantially alter the thermal,
chemical, ionization, and dynamical (e.g., laminar $vs.$
turbulent flow) state of the disk.  Long-standing mysteries
in solar nebula studies, such as the flash melting of
meteoritic chondrules and the presence of short-lived
radionuclides, may be explained, at least in part, by these
high energy phenomena in young stellar systems.  Planet
formation and early evolution such as inward migration
should be substantially different in an irradiated disk as
compared to an isolated disk.

Disk solids will be affected both by the observed X-rays and by other
high energy manifestations of the violent magnetic connection
process that heats the X-ray emitting plasma.  These flare events
should produce shock fronts analogous to solar coronal mass
ejections (CMEs) accompanied by baryons and electrons accelerated
to MeV energies.  For example, the solar X17 flare of 28 Oct 2003
produced a peak fluence of $8 \times 10^5$ protons cm$^{-2}$
s$^{-1}$ ster$^{-1}$ MeV$^{-1}$ around 1 keV and $8 \times 10^4$
protons cm$^{-2}$ s$^{-1}$ ster$^{-1}$ MeV$^{-1}$ around 1 MeV, as
measured by the Advanced Composition Explorer.  This represents a rise of a
factor of $10^4$ in particle fluence from the quiescent
level and was accompanied by a $10^3$-fold outburst in X-ray
emission with peak X-ray luminosity $L_x \simeq 4 \times 10^{27}$ erg
s$^{-1}$ seen with the GOES\,12 satellite. When scaled upward to
the $10^{28}-10^{32}$ ergs s$^{-1}$ flares seen in T Tauri stars
surrounded by protoplanetary disks, the inferred X-ray or shock
wave heating may flash melt chondrules (Shu \e 2001, Nakamoto \e 2005) and the
inferred MeV particle fluence may produce anomalous abundances of
short-lived radionuclides seen in Ca-Al-rich inclusions (CAIs) of
ancient carbonaceous chondritic meteorites (Feigelson 1982,
Gounelle et al. 2001,  Goswami, Marhas \& Sahijpal 2001, 
Leya, Halliday \& Wieler 2003).

The study of the effects of high energy products of magnetic
flares on young circumstellar disks is being propelled by
observations with the $Chandra$ $X-ray$ $Observatory$
with its large collecting area and
elliptic orbit that permits long continuous exposures.  
The VLA radio telescope also provides direct measurements 
of gyrosynchrotron emission from MeV electrons in some young stars (G\"udel 2002). 
Feigelson, Garmire, \& Pravdo (2002; henceforth FGP02) investigated
the X-ray flaring of a complete sample of 43 solar-mass stars in the
Orion Nebula Cluster (ONC) observed with $Chandra$'s Advanced CCD
Imaging Spectrometer (ACIS) in two $\sim 12-$hour exposures
during 1999-2000.  They found that solar analogs with ages $\leq
1$\,Myr exhibit flares $\simeq 30$ times more powerful and $\simeq
300$ times more often than the most powerful flares
on the contemporary Sun and argued that such flares
imply factors of $10^5$ enhancement in the MeV proton fluence from
the young Sun. They conclude that this is sufficient to explain some of
the important CAI isotopic anomalies  by {\it in situ}
  spallation reactions.

The present paper continues this effort to measure
flaring properties of young solar analogs directly and understand the impact on
protoplanetary disks. 
The principal goal of this study is to obtain quantitative measures of the 
variability, particularly the impulsive flaring, of pre-main sequence analogs 
of our Sun.  We establish the frequency, duration and energetics of flaring, 
and strive to understand their effects on the protoplanetary disks that 
surround many of these stars.   We then study the effect of the
flares on the disk.  Our paper is based on a
nearly continuous observation of the ONC made with $Chandra$'s
ACIS detector over a 13.2-day period in January 2003.  Known as
the $Chandra$ Orion Ultradeep Project (COUP), this observation
provides the most comprehensive database for studies of young
stellar X-rays and magnetic activity (Getman \e 2005; henceforth
G05). 

Due to strong dependencies of X-ray luminosity on
stellar mass (Preibisch \e 2005a), we limit this study to stars
with masses close to 1.0 M$_\odot$.
Section 2 defines the sample, \S 3 describes our extraction of
flares from the complex X-ray light curves, and \S 4 outlines our
spectral analysis procedures. Results on the X-ray intensities,
durations, frequencies, energetics and dependencies on stellar age
and disk properties are presented in \S 5.  Implications for
effects of protoplanetary disk gases and solids are outlined in \S
6.

\section{Selection of Sources}

The COUP observation, field of view, X-ray source detection, event
extraction, X-ray properties and stellar counterparts are
described in detail by G05, who give extensive tables of source
properties.  The present paper gives a close examination of the
subset of COUP sources with masses $0.9 < M < 1.2$ M$_\odot$,
which is roughly in the mass range of F7-G5 stars on the Zero
Age Main Sequence (ZAMS).  This selection permits comparison with samples of
ZAMS G stars from the Pleiades, Hyades and other stellar samples
(Preibisch 2005b).

Twenty-eight COUP sources in Table~9 of G05 and two undetected stars in their
Table~11 have masses in the desired range giving a total
sample size of 30 stars for our study. These sources are shown on
the COUP and 2MASS fields in Figure 1.  These masses were derived
in G05 by applying the evolutionary tracks of 
Siess, Dufour, \& Forestini (2000; henceforth
SDF00) to stellar positions in the Hertzsprung-Russell
diagram derived from optical spectroscopy and photometry of $V<20$
ONC stars by Hillenbrand (1997) [as updated in G05].  Choices of mass
tracks and spectral type to temperature conversions used in G05 were
made to maximize utility and were not optimized for a particular
study.  The sample used here differs from that 
obtained by FGP02, who also used the HR Diagram 
of Hillenbrand (1997) but applied the theoretical tracks of
D'Antona \& Mazzitelli (1997; henceforth DM97). The SDF00 calculations have
different treatments of degenerate electron pressure, chemical
composition and mixing length parameter from those of DM97 tracks.
While the two sets tracks show similar performance with respect to
dynamically measured masses (Hillenbrand \& White 2004), the SDF00
tracks were selected in G05 for their applicability to a wider
range of masses for the full COUP study. 

How well the stars selected here truly represent ``young Suns'' depends
strongly on the ability of models to predict accurate masses from
temperatures and luminosities and upon limitations in measuring those quantities. 
Compared to the DM97 tracks, the SDF00 tracks predict
surface temperatures up to 600 K cooler and higher
luminosities for
$\simeq 1$ M$_{\odot}$ stars along the convective Hayashi tracks.
Consequently, for a given luminosity the SDF00 age estimate is
younger and the mass estimate is higher.  The masses range from
10\% to 100\% increases over those predicted by DM97.  The result
is that most of the 30 stars considered here lie below the
$0.8-1.4$ M$_\odot$ range considered by FGP02 -- only 7 stars are
in common between the two samples.  The samples would have more closely
overlapped had we considered the $1.2 < M_\odot < 1.4$ M$_\odot$ SDF00 mass
range.  

Another selection effect implicit in the use of the Hillenbrand (1997) 
spectroscopic sample is the omission of
heavily obscured ONC stars.  The highest extinction in our
subsample is $A_V \approx 6.3$.  Thus, we exclude a number of obscured
stars whose location in the $K-(H-K)$ diagram
(see e.g., Hillenbrand \& Carpenter 2000) predicts a mass around 1\,M$_\odot$.
We also omit the weak-lined T Tauri star GMR-A with $A_V \simeq
35$ and an inferred mass slightly over 1 M$_\odot$; it exhibited
extraordinarily powerful radio and X-ray flares during the COUP
observation (Bower \e  2003). Again, we believe that the omission
of high-$A_V$ stars from consideration will have little or no
effect on our conclusions regarding magnetic activity of solar
analogs.  It has the further advantage that soft X-ray absorption
in our sample is never very strong with column densities $\log N_H
\leq 22.0$ cm$^{-2}$.

Table 1 lists measured optical and near-infrared properties of the
sample stars, and Table~2 gives inferred properties such as effective temperature
$\log T_{eff}$, bolometric luminosity $\log L_{bol}$, radius $R$,
mass $M$, age $t$, in addition to the  $K$-band excess ($\Delta (I-K)$) and
the equivalent width of the \ion{Ca}{2} triplet. 
We do not tabulate the errors in the table for clarity, but they can
be significant. A recent analysis by 
Hillenbrand \& White (2004) found that PMS stars of
about 1M$_\odot$ have luminosity errors of about 0.1 dex.  This reflects 
typical spectral subclass errors (of one subclass), photometric errors
introduced both by observational error (less than 3\%),
 variability and extinction (errors of $<$ 10\%).
The choice of a particular PMS evolutionary mass model
will introduce a bias in source selection.  Hillenbrand \& White
found that the DM97 and SDF00 models both 
systematically underestimate sub-solar masses by 10\%--30\% on average,
with scatter of the same order.  Further, we note the very 
non--uniform mass distribution among the sources listed in Table~2:
only one source lies between 0.95 and 1.09 M$_\odot$.  This is
caused by the quantization of temperature to 200 degree increments
(0.02 dex in log T) by G05.  A temperature difference of 200K corresponds 
to 0.2 M$_\odot$ along the convective part of the evolutionary track in
these models. This limitation of the dataset quantizes sample masses
somewhat and results in a systematic shift in the attributed mass of
stars, with masses between 1.3--1.0 M$_\odot$ shifting toward 1.12 M$_\odot$,
and masses between 1.0--0.8 M$_\odot$ shifting toward 0.90 M$_\odot$.
We cannot limit the mass selection as tightly
as we had hoped for stars younger than about log 6.5 yrs, but
we believe these issues have little effect on our
conclusions, other than shifting slightly downward the
average X-ray luminosities as compared to FGP02 
(see also Preibisch 2005a).

All quantities in Table~1 are obtained from the tables of G05 
where a full description of their derivation is given.  
Throughout this paper, we refer
to sources by their COUP number which ranges from $1-1616$.  The
corresponding optical identifier is the star number in
Jones and Walker (1988).  There is virtually no 
uncertainty in the association of COUP and
bright optical ONC stars as the astrometric offsets are typically
$<0.2$\arcsec~ (Table 1, columns 3-4), although contributions to
both optical and X-ray flux by lower mass close binary companions
may be present.  In two cases, the optical field shows a visual
double where only the brighter component is associated with the
COUP source. 

Table 3 gives time-averaged X-ray luminosities selected from Tables
4 and 8 of G05.  Column 2 gives the number of extracted counts (NetCts) from
the position-dependent source region, after background-correction.  
Columns 3$-$6 give time-averaged luminosities in three bands --
$\log L_s$ in the soft $0.5-2$ keV band, $\log L_h$ in the hard
$2-8$ keV band, and $\log L_t$ in the total $0.5-8$ keV band --
and the absorption-corrected intrinsic total band luminosity $\log
L_{t,c}$.  The final column lists notes about each source.

Note that the two solar analog stars that were
{\bf not} detected by the COUP (from Table 11 in G05), JW 252 and JW
407, are
both close to other bright COUP sources which raised the local
background.
JW 252 has estimated mass of $1.18$ M$_\odot$ and an age of $2.5$ Myr 
but lies in the wings of COUP\,222 associated with JW 256 lying  3.8\arcsec\/ to the NE. 
JW 407 has estimated mass of $0.91$ M$_\odot$ and an age of $2.0$ Myr 
but lies 3.4\arcsec\/ to the ESE of and in the wings of COUP\,520.  COUP\,520 is not associated
with an optical source but is associated with 2MASS 05351317-0517307 with an offset of 
$< 0.15$\arcsec.  These stars  are probably not qualitatively different from
the resolved sources, and will not be further discussed in this
paper.

In all but four cases, the optically-derived visual absorptions
$A_V$ agree with the X-ray-derived column densities $\log N_H$
given the conversion $N_H = 1.6 \times 10^{21}~A_V$ cm$^{-2}$
obtained by Vuong \e (2003).  COUP\,241, 314 and 1167 are
discrepant in that they exhibit almost no counts below 0.8 keV and
thus have high inferred $\log N_H$ value, inconsistent with the
low reported $A_V$.  COUP\,1539 is discrepant in the other
direction possibly due to an unusually soft X-ray flare.

We estimate that about half of the COUP-detected solar mass ONC
stars possess circumstellar disks based on three criteria.  First,
three stars have proplyds, disks imaged in silhouette against the
bright nebula or imaged in H$\alpha$ emission by the $Hubble$
$Space$ $Telescope$.  An additional 13 stars appear to have inner
dusty disks producing $K$-band excesses $\Delta(I-K) > 0.3$ over
the emission expected from an isolated photosphere.  Third, three
of these sixteen stars show significant \ion{Ca}{2} triplet
emission, EW(\ion{Ca}{2})$\leq -3$\AA, which is a rough indicator of
accretion (Flaccomio \e 2003, Sicilia-Aguilar \e 2005).

\section{Definition of a Flare}
The distinguishing aspect of the COUP dataset is the unprecedented
duration of nearly continuous observation.  This allows a unique
perspective of the temporal behavior of the X-ray sources.
While flaring is
ubiquitous among coronal X-ray sources,
the temporal behavior of X-ray sources is complex.
Separating obvious
flares and less obvious periods of variable emission, determining
``quiescent'' flux levels and quantifying the intensities of flaring have all been 
challenges. 

\subsection {Quantifying  Variability}
Several groups have used different techniques to quantify variability in
stellar X-ray sources. Observations of stars by ROSAT were typically
divided into fairly short observation intervals (OBIs) of order 3 ks.
This prevented observation of flares from beginning to end but  
allowed observers to easily compare count rates among various
OBIs and often define flares as the doubling of flux in one OBI
as compared to another. A more rigorous
technique commonly used is a one-sample Kolmogorov-Smirnov (KS) test,
but virtually every X-ray source (star or otherwise)  
varies if observed for a sufficient duration
and with sufficient sensitivity.  G05 reports that 974 of
1616 X-ray sources in the COUP data set were non-constant at a 
confidence level of greater than 99\%.  The solar mass stars
discussed here are all non-constant at a confidence of 99\%. 

There is a plethora of other methods that have been used to
determine variability.
These methods include: distribution of binned count rate (Saar \& Bookbinder 1998, G\"udel 2003),
 n -- $\sigma$ deviation from the mean (Stelzer \e 2000, Wolk \e 2004),   
Lagrange multipliers (Schwartz 1987), Poisson tests
(e.g., Maccacaro \e 1987), avalanche and cellular automata models
(Lu \& Hamilton 1991, Lu 1995), phase space reconstruction
(Vio \e 1992), KS visualization (Giommi \e 1995),
wavelet analysis (Aschwanden \e 1998; Walker \e 2000; many others), 
Poissonian structure function 
(Fernandes \e  2000), time-frequency analysis (Vio \& Wamsteker 2002)  and
combined K-S/$\chi^2$ criteria (Fuhrmeister \& Schmitt 2003). We sought
a method that is not only independent of data binning but also
identifies blocks of data in a specific, not statistical sense.

We have chosen to employ a method to determine periods of constant signal 
similar to the Bayesian Block method discussed by Scargle (1998) and
used in G05. 
The Bayesian Block 
technique segments the lightcurve into a sequence of constant 
brightness levels under the assumption of Poisson errors in the signal.  
The change points between levels are established using an iterative 
maximum likelihood procedure based on Bayesian principles.  
This method is explicitly designed to avoid binning of 
the observation into equally spaced time intervals.  
Bayesian Blocks do characterize the X-ray
lightcurves remarkably well.  As discussed in G05, the overall
shape and number of  Bayesian Blocks is fairly insensitive to the
choice of confidence level.  

 We adopt here a procedure very similar to Bayesian Blocks but without 
use of Bayes' Theorem.  The sequence of constant brightness levels is
established by maximizing likelihoods under the Poisson model in the 
same way, but our thresholds for establishing change points seek to minimize 
false positives to reduce fragmentation of Blocks into many small intervals.  
These thresholds were established from extensive simulations of constant 
lightcurves at different count rates.  
We call our procedure, used both 
here and in other COUP studies, Maximum Likelihood Blocks (MLB).

Further, we have made an attempt to overcome
one of the main limitations of the Scargle method --  the fact that
it segments a light curve in only two segments at a time (the
algorithm is then run recursively on each of the segments found). This
is a limitation especially for finding faint impulsive events (e.g., flares),
as a two segment representation in which one segment includes the event
might not be statistically significant 
so that the segmentation process cannot start. 
For this reason, our code tests both the two- and the three-segment
hypotheses. For  computational efficiency, only segments with 
$<2000$ counts are tested for the 3-segment hypothesis.

We found that using one count
per average segment occasionally produced too many short blocks. 
For analysis in this study, the two parameters of the MLB algorithm were set to require 
a minimum of 20 counts per block and a 95\% probability for establishing change points.  
For other COUP studies (e.g., involving much fainter sources), other parameters may be chosen.  
The median\footnote{In this paper, we will generally quote the median not
the mean since the median 
is insulated from outliers that can be the result of poor fits or 
low counts. When a dispersion is needed, we report the median absolute deviation 
(MAD; Beers \e 1990).}   
observation in our sample contains 12 such blocks, five being the smallest
number of blocks and 31 being the largest.  The existence of multiple 
blocks tells us that the flux rate from all sources is non-constant
at high confidence, but the number of blocks alone does not tell us
anything about the type or level of activity detected.

\subsection{The Morphology of Stellar X-ray Variability}

Solar and stellar activity is often characterized in terms of
``quiescent'' versus ``flaring'' 
state.   The latter state is sometimes divided into macroscopic distinct events and 
the superposition of many microflares or nanoflares.  We consider only 
macroscopic flaring in this study.  Although many COUP lightcurves
exhibit frequent long roughly-constant periods of weaker emission between powerful 
flares, the term ``quiescent'' (dictionary definition: ``marked by 
inactivity'') is a misnomer because the absolute activity level is often 
extremely high; e.g., $\log (L_t/L_{bol}) \simeq -4$ in contrast to -6 for 
the quiet Sun.  We thus chose the term ``characteristic'' level to describe 
the typical emission between isolated flare events from a COUP star.
We term the single flux block of weakest emission as the
``Minimum Observed LEvel'' (MOLE).

Upon subjective examination, the blocks divided themselves into three types:
I) The {\bf characteristic} level and blocks statistically compatible with the
characteristic level. 
II) Periods of flux ``elevated'' above the characteristic level with no
impulsive  morphology.
III) Periods of {\bf elevated} flux marked by very rapid rises. 
We refer to only the third group as ``{\bf flares};'' this was the only group 
exhibiting consistent morphology.   Specifically, all
the events in group III showed either ``classic'' flare morphology
(i.e., a fast rise followed by an exponential decay) or a symmetric
morphology similar to that seen during EUVE observations of HR 1099 
(Osten \& Brown 1999).  We also found it useful to define an 
intermediate ``{\bf very elevated}" level for unusually strong emission without rapid flux changes.

The classification of X-ray variations from our MLB segmentation of the COUP 
lightcurves is illustrated in Figure~\ref{sample} for COUP\,567.
The top panel shows the binned lightcurve overlaid with the 25 MLBs used in our analysis.

\begin{description}

\item{Characteristic Level}~~ 
-- We established the characteristic level iteratively, first identifying
the set of blocks that maximize the period of time covered
by compatible blocks.  
A segment is deemed compatible with count rate, $R$,
if its count rate, $R_{block}\pm \sigma$, is between  R/1.2-1.5$\sigma$ and
R*1.2+1.5$\sigma.$  Here, $\sigma$ is the standard deviation of the
observed rate about the mean rate in the block.
A reference count rate  $R_{ref}$
is defined as that
rate for which the compatible time reaches a maximum. 
Then characteristic segments are re--selected to include
\begin{equation}
R_{block} <  1.2 \times R_{char}+1.5\sigma 
\end{equation}
Although there are cases in which it is very
difficult, even for the eye, to define a characteristic level, this algorithm
gives reasonable results.    In the middle panel of
Figure~\ref{sample} there are 11 characteristic MLBs.

\item {Elevated Level}~~ -- Elevated periods are defined to be 
slightly elevated and not associated with macroscopic flaring events.
\begin{equation}
 2.5 \times R_{char}+1.5\sigma > R_{block} > 1.2 \times R_{char}+1.5\sigma
\end{equation}

In the middle panel of Figure~\ref{sample} there are nine such blocks,
only one of which is not associated with a flare.

\item {Very Elevated Level}~~ --  The remaining blocks are defined as very 
elevated periods:

\begin{equation}
R_{block} >  2.5 \times  R_{char}+1.5\sigma
\end{equation}

These are intense enough that they are often associated 
with macroscopic flaring events. In Figure~\ref{sample}, there are five such blocks.

\item {Flaring}~~ --  The key to identifying a flare by 
eye within a
``very elevated'' period is its rapid flux change.  Hence, 
it is natural to add the derivative of the lightcurve (dR/dt)  
(or the second derivative of the photon arrival time) to the criteria.  
The difference between successive block rates can easily be chosen as dR, but defining dt 
requires careful thought. A simple definition of dt as the interval
between the mid-time of consecutive blocks would dilute truly rapid
rises, especially when observation gaps exist.  We instead choose
dt as the shorter of the exposure times in two successive blocks.
To prevent a dependence on rate, 
dR/dt is scaled by the inverse of $R_{char}$.
By plotting  $1/R_{char}\times dR/dt$ and the lightcurve 
simultaneously we empirically determined a
threshold of $1/R_{char}\times dR/dt > 10^{-4}$ s$^{-1}$ to be indicative of a
flare.  A flare is defined as a successive series of elevated blocks which
include at least one very elevated block and one period in which 
\begin{equation}
1/R_{char}\times dR/dt > 10^{-4} ~~\rm{s}^{-1} 
\end{equation}
In the bottom panel of Figure~\ref{sample}, we plot 24 values of:
\begin{equation}
\Delta= 1/R_{char}\times dR/dt
\end{equation}

Ten of these exceed our threshold, and seven are
associated with very elevated levels.  Grouping the consecutive
occurrences, we find three flares.    The blue lines in the third panel
of Figure~\ref{sample} and the similar panels of Figure~3 identify 
the full durations of the flares.
\end{description}


A weakness of this set of definitions is
the rather stringent requirement of flare strength.
Only intervals for which the stellar luminosity rises by more than a
factor   $2.5\sigma$ above 120\% of the characteristic level are 
definable as flares.  In practice, the smallest luminosity change
associated with a flare is a luminosity change of a factor of three.
Flares moderately weaker than this are relegated to the
``elevated'' category.  Any activity that increases the overall flux
by less than $\sim$ 50\% is completely ignored and accounted for in
the characteristic data.

The levels chosen for our definitions are admittedly subjective -- especially 
the factors 1.2 and 2.5$\sigma$ that enter in the definition of elevated vs.
characteristic levels and the setting of the flare level  
$\Delta$ at $1/R_{char}\times dR/dt > 10^{-4}$ s$^{-1}$.  These definitions 
are tuned to match what we intuitively agree are significant flares.
Some probable flares are missed, such as the second very elevated
period for COUP\,1134.\footnote{Since it is seen just coming out of a perigee
passage, its rise time is seriously overestimated and its slow decay
prevents detection of a rapid decline.} Some marginal events are also counted
such as flares, e.g., the final three flares on COUP\,314, where a low characteristic rate
serves to exaggerate the rate of change.
In testing several variations on these
numbers to arrive at our final criteria, we found that some weaker
flares are lost and others are found.  Based on this experimentation, we estimate that the total number of 
flares is reliable at the $\pm10$\% level with respect to variations in the criteria of
equations (1)--(4).

Figure~\ref{lc14}, shows the lightcurves for the 28
solar mass sources in the ONC. The lightcurves are presented as
histograms with blocks of constant flux overlaid.  A new block is used
when maximum likelihood statistics indicate 95\% confidence
that the data are not consistent with a constant signal.  
The lower plot in each panel indicates $1/R_{char}\times dR/dt$ between
each pair of blocks.  Flares are indicated with lines above the lightcurves.

\section{Spectral Fitting}

Much of the analysis of pre-main sequence flaring in this study will 
be based on spectral modeling of the COUP source spectra in terms 
of optically-thin plasmas in collisional equilibrium.  From
this modeling we extract plasma temperatures, instantaneous luminosities 
and time-integrated energy output in the $Chandra$ $0.5-8$ keV band.  
We assume elemental abundances follow the cosmic abundance 
pattern with 0.3 times the solar abundance.  This may be a poor 
assumption as abundance anomalies and temporal variations have 
been seen in other magnetically active stars (e.g., Favata \& Schmitt 
1999, Brinkman \e 2001) and are present in some COUP sources 
(G05).  Abundance effects will be treated in a later COUP study, 
but their omission here should have little effect on our
determinations of broad-band luminosities and energies.

We employ the usual corrections as discussed in G05 to
fit the spectra of the photons which arrived during characteristic
periods to a two-temperature MeKaL plasma.   
MeKaL was used as a convention by the COUP team.
We found that if \nh is left as a free parameter, 
there is a strong correlation between luminosity and \nh 
(luminosity increasing with increasing \NH).   This is due to a natural
degeneracy in the problem:  one can arrive at a uniformly good fit
by increasing \nh and the high temperature flux in parallel.\footnote{This
degeneracy is exacerbated when using the front illuminated CCDs of
ACIS-I since they have  
low effective area at low energy where the absorption is most important.}
For this reason, \nh was frozen at $A_v\times 1.6 \times 10^{21}$.
An exception was made when the fit was formally  
poor; when $\chi^2/d.o.f > 2.5$, we allowed \nh to become a free parameter.

Once the characteristic spectra are determined, we proceed
with the remaining spectral analysis of each star.  
All photons that arrive
during elevated periods are combined into a composite elevated spectrum for each
star.   Photons that arrive during flare periods are separated 
to create spectra for each flare.  
Except where noted, each of these $\sim$ 70 
spectra is fit by a one or two-temperature MeKaL plasma with \nh and
metallicity frozen to the characteristic values.
We show an example of this fitting process for COUP\,262 in  Figure~\ref{sampleq}.
In practice, four stars\footnote{COUP\,241, 250, 262 and 314.} 
needed \nh to become a free parameter
regardless of whether characteristic, elevated or flare data were being fit.
This could be due to high local extinction, 
to unusually high coronal temperatures, to a broad range of coronal
temperatures or to poor extinction estimates.   Data for these
stars should be treated cautiously.
In addition, no spectral fits were performed for COUP\,828 in flare because its
proximity to a chip gap confused our flare detection algorithm.

\section{Analysis}

In this section, we examine the sources in the three states,
characteristic, elevated and flare.  We compare the luminosity and
plasma temperatures in these states.  We focus characteristics which
will have the greatest effect on matter near the star, including the peak luminosity and
temperature, as well as the frequency and duration of flares.

\subsection{The Characteristic Level}

Table~\ref{char_table} 
shows the results of  the spectral fits for each star's characteristic
level.  The first column indicates the COUP source ID, the second
gives the interval the source spent at the characteristic level.
Columns 3--5 give the number of bins, the $\chi^2$ per degree of
freedom and the null probability.  Columns 6--11 give \nh 
(converted from A$_V$ or fit), the
temperatures of the cool and hot coronal components, ratios
of the emission measures of the cool to hot component,
the resultant luminosity assuming a distance of 450~pc and the energy
released between 500~ev and 8~keV during the indicated interval.
The final column notes where \nh was allowed to be a free parameter.

 On average, sources were at their characteristic levels for about 
640~ks out of 850~ks or 75\% of the time.\footnote{This is in 
remarkable agreement with Osten \e
(2004). Over a four year coordinated campaign, they found that the RS CVn
System  HR~1099 has a 30\% duty
cycle for coherent low frequency radio emission.  The flare frequency
cited by Osten \& Brown (1999) is somewhat higher ($\sim 40 \%$).}  
Restricting our sample to sources with high-quality 
(i.e., acceptable $\chi^2$) two-temperature fits, the median luminosity 
of the characteristic level is $\log L_{t} = 30.25$ ergs s$^{-1}$. 
The dashed line in Figure~\ref{lum_hist} shows the distribution of
these characteristic luminosities.
This is consistent with the results from Flaccomio \e (2003) who
found a median luminosity of  $\log L_{t} \sim 30.5$ ergs s$^{-1}$ for stars between 1
and 2 M$_{\odot}$ and  $\log L_{t} \sim 30.05$ ergs s$^{-1}$ for stars between 
0.5 and 1 M$_{\odot}$.  Their use of a ``basal flux'' (similar to our
characteristic flux)  mitigated bias due 
to large flares that occurred on about 5-10\% of the stars in their
sample.  Our result is slightly lower than the average luminosity found
for solar mass stars by  FGP02.
The characteristic $\log L_{t,c}$ is roughly uniformly distributed 
with a MAD of about half a dex (0.49).

We expected to find L$_{char}$ and
bolometric luminosity to be well correlated since the flaring
component has been removed from the determination of X-ray luminosity. Indeed, 
both Spearman ($\tau$) and Kendall ($\rho$)
correlation coefficients show greater than 99.98\% confidence in the correlation.
A two-variable linear regression first order fit finds 
$\log L_{char} \propto \log L_{bol}$ such that 
$\log L_{char}/ \log L_{bol} \simeq -3.58$ with $MAD = 0.21$.
In these data, we focus solely on 
the characteristic X-ray flux from the star.
In summary, about 0.03\% of a star's 
luminosity is characteristically released as X-rays between 500~eV and 8~KeV.

We also examined the relation between characteristic luminosity and
age. We find  $\rho =$ 0.12, a weaker correlation than that reported by FGP02 
and consistent with no correlation.
As shown in Figure~\ref{lt_age}, we fit the data to an outlier-resistant 
two-variable linear regression giving 
 log $L_{char} =32.1-0.47 {\rm log} (t)$,  
where t is in units of a million years.  
The $1\sigma$ scatter from this fit 
is 0.35 in units of $\log L_{t,c}$.
This is a flatter relationship than that reported by FGP02 who found a
slope of $1.1 \log(t)$.  Our slower evolution within the COUP age range
was also found in the COUP study of 
Preibisch \& Feigelson (2005).

\subsection{Elevated and Very Elevated Levels}

FGP02  used ``a simple subjective classification 
of the variations" observed in solar mass stars.
They classified stars as constant, long term variable, possible flare
and flare.
They found $\sim$ 30 flares or possible flares on 43 stars (69\%) 
in 50 ks of  observing time. 
At that rate,  we should see $\sim$ 12 flares per star or about 
350 total flares, and in fact, this is very close to our total number of 
MLBs.
Using our definitions, we find $\sim$ 40 flares on 28 stars in about
17 times the observing time of the FGP02 data.  
In this sense, our definition is about a factor 10 times more 
conservative than that of FGP02.
We have instead termed these $\sim$ 300 ``missing'' flares as
``elevated" level
intervals; this activity is not ``missed'', it is merely accounted differently.  
Since we are concerned with the effect on 
the protoplanetary disks, the relevant quantities are the
duration, luminosity and temperature of X-rays during elevated periods.

Table~\ref{tbl_e1}
shows the results of  the spectral fits to data from each star's
periods of elevated flux.  The column definitions are the same as
those for Table~~\ref{char_table}.
The median duration in the elevated regime
is about 65~ks out of the total 850~ks; i.e., on average, sources are 
elevated only about 8\% of the time. This is consistent with the
definition that the flux is between 1.5 and 2.5 $\sigma$ above the
characteristic level, and
luminosity changes are similarly limited to be between 1.6 and 2.6 
times the characteristic luminosity.  There is evidence that the
periods of elevated flux are indeed associated with microflaring as
the median temperature of the hot component of the corona goes up by
25\% during these periods - from a median of 2.34 keV (MAD = 0.7) to a median of
2.93 keV (MAD = 0.7) - while the cool component stays constant.

There are only 6 periods of very elevated rates that failed to  
start with a rapid rise of $\Delta>10^{-4}$ and were therefore 
not classified as flares. 
The duration of non-flaring yet very elevated
fluxes represents only 1.2\% of the total observing time.  The very elevated
periods are associated with modest luminosity changes of between 2.5 and 5.
The median temperature of the hot component of the corona goes up 
further to about 3.1 keV while the temperature of the 
cool component stays constant at around 800 eV.

\subsection{Flaring}

\subsubsection {Strength and Duration of Flares \label{stre_dur}} 
Forty-one flares were detected at 95\% confidence.\footnote{Using 
maximum likelihood statistics, blocks are
formed on the basis of confidence
that the data are not consistent with constant and hence a new block
is needed. The determination of a flare follows equations (1)-(4).
The determination of the duration and the rates of the
blocks are subject to the confidence in that block.}
There is no single objective method for 
determining the duration of a flare, but there
are two obvious extremes: concentrating just on the temporal region of
the peak of the flare or considering the entire time that the count
rate is elevated.  
In adapting the latter ``whole flare'' approach, we looked at
``prototypical'' flares like that of COUP\,262 
or the two flares on COUP\,223.  
In these cases, the tail of the flare includes both ``elevated''
and ``very elevated'' blocks.   Here we could define the duration of a flare
as the total duration of all successive elevated blocks adjacent to a
flare. The range of flare durations is vast, from 1 hr to 3 days  as shown in
Figure~\ref{duration}.

Table~\ref{tbl-flares} contains the spectral fits to all the
flares with at least 50 photons to fit.  The columns in this table are
identical to the two previous tables except that the results for each 
flare are tabulated separately.  The luminosities and
temperatures represent averages of each flare whereas the energy gives the
total released in the event.
The median $\log L_{t,c}$ was about 30.81 ergs s$^{-1}$ with the median 
temperature of the hot coronae of about 3.45~keV.  The cool 
corona remained near 800~eV.  
The ratios of flare to characteristic luminosities, plotted in
Figure~\ref{ratio},  show that most flare amplitudes are not 
extremely high.  The median flare level is 3.5 times the
characteristic level, and 90\% of flares rise to less than 
10 times the characteristic level.  

The luminosities of the
  peak blocks are another measure of flare intensity.
Table~\ref{tbl-brightest} lists the spectral fits to just the peak of
each flare (if there were enough photons to yield a reliable fit).
As can be inferred from  
Figure~\ref{lum_hist}, the peak flare luminosities are about 0.25
dex higher than the integrated flare 
luminosities (median $\log L_{t,c}/ \log L_{bol}$ 
=-2.77).  This method of calculation has little effect on the 
relative strength deduced for each flare, the median ratio of the
peak to characteristic luminosity is 6.
The largest change was
observed during the second flare of COUP\,1259, when the ratio of the peak to 
characteristic luminosity soared to 80.
The median peak $\log L_{t,c}$, was
about 30.97 ergs s$^{-1}$, while the 
median duration of peak blocks is about 10~ks.  More importantly,
the median plasma temperature was about 7 keV during the peak of each flare!  
The most luminous period is also
likely the hottest portion of the flare and is therefore 
the most capable of affecting circumstellar material.

There are no strong trends of the duration of flares relating to the
luminosity. Specifically we find no direct relation between flare
duration and either peak intensity or average luminosity. 
The $\rho$ rank correlations were 0.88 and 0.41 
respectively -- consistent with no correlation.
We find a median ``whole flare'' duration of
about 65 ks or about 6.5 time longer than the period of peak intensity.

\subsubsection {Energy Release in Flares}

Having determined the duration and the luminosity of each flare we 
can now discuss the energy released during each event.
Multiplying the full duration of each flare by its luminosity
(columns 3 and 11 of Table~\ref{tbl-flares}) we calculated the total energy
released during the event (column 12):
The median flare released about 10$^{35.5}$ ergs.
Most energetic, was the
second flare on COUP\,1259 which was an order of magnitude brighter than 
any other.  Figure~\ref{energy} shows the cumulative 
number of flares for a given energy.  Fitting the data below $10^{36}$
ergs demonstrates that the number of flares releasing a certain energy can
be expressed as N= 1.1 $\log E^{-0.66}$. Differentiating the equation
solves for the number of flares as a function of energy:  $dN/dE
\propto E^{\alpha}$ with
$\alpha  \simeq 1.7$. The largest uncertainties in the fit are not
statistical, but rather
come from the fitted range.  Constraining the fit to energies below
$10^{35}$ ergs the power law index is decreased by 0.11 to $\alpha=1.55$.

Measurements of the X-ray flare energy distribution index lie
in the range $1.5 < \alpha < 2.5$ for the Sun and $1.5 < \alpha <
2.7$ for magnetically active stars (G\"udel et al.\ 2003).
A key difference for the Sun is that the total energy released in a flare 
is less than  $10^{32}$ \erg\ (cf. Figure~6 in Crosby, Aschwanden \& Dennis
1993).
Several groups  (Collura \e 1988,
Pallavicini, Tagliaferri, \& Stella 1990) 
derived $\alpha$ = 1.5-1.7 for a sample of M-dwarf flare observations
with EXOSAT.  Our result is also consistent with 
distribution of X-ray flares on the contemporary sun and other active stars
reported by earlier authors $dN/dE
\propto E^{-1.7\pm 0.1}$ (Hudson 1991, Crosby, Aschwanden \& Dennis
1993).  Recent statistical investigations suggest $\alpha$ =
2.0--2.6 for small flares in the quiet solar corona 
(Krucker \& Benz 1998; Parnell \& Jupp 2000).

Our flare distribution is cut off below 10$^{34}$ ergs which arises from the
  classification criteria that require the
flares to lie $\geq 3$ times above the characteristic luminosities 
around $\log L_{t} \simeq 30.5$ ergs s$^{-1}$.  
The overall lightcurves are not consistent with all the 
X--rays being produced by a single power law extending 
down to zero counts. The overall appearance of the
lightcurves is a relatively constant characteristic level with
sudden big flares.  Inspection does not show obvious low frequency signal
which would be expected  
due to the superposition of intermediate and low-intensity flares as implied by a power law.

To demonstrate this in a qualitative manner
with a simple simulation.  We created a simulated 13.2 day
lightcurve with a ``toy'' template flare (rapid rise, 10-hr exponential
decay and  a distribution following a power law. The main free
parameter was dN/dC, 
where C is the number of counts in a flare (modeling the energy is
beyond the scope of this study).  
Simulations were performed 
with $dN/dC\propto C^{-1.7}$, $dN/dC \propto C^{-2}$, 
$dN/dC \propto C^{-2.5}$ and
$dN/dC\propto C^{-3}$,  and created lightcurves by the superposition of flares.
To compare results for high and low count rate sources, 
the number of flares was varied from  100 to 4400.  For each power law index 
100 simulations were run at each flare quantity. 

Some examples of our 43,000 simulations are shown in Figure \ref{sims}.
By inspection, it appears that if all of the X-rays are produced by
flaring and if the flares follow a single power law, values of the
power law index of -1.7 would create more flares than are seen in our data.
Similarly, the power law index of -3 produces too few flares. 
If the sole source of the X-rays were flaring drawn on a single
power law, the index would have to about -2.25, which is too
steep to describe the brightest flares.  Thus, either flare energies
are better represented by a broken power law, or by a single power law
overlying 
an X-ray continuum.  G\"udel \e (2003) recently reported on
EUVE and BeppoSAX observation of the nearby dMe star AD Leo, 
ascribing all the activity to small--scale flaring with $\alpha$ 
values between 2.0 and 2.5.
A full quantitative comparison with the whole COUP 
dataset is beyond the scope of this paper and would require
more extensive work over a broad range of masses.

The total energy release in each flare is nearly linearly correlated with
the luminosity.  The best fit to a linear regression is
\begin{equation}
Energy \propto Luminosity^{1.16}
\end{equation}
The MAD of the fit is 0.38 dex and the
rank correlation is correspondingly strong ($< 10^{-5}$).
Since energy and luminosity are linearly related \ie\
  Energy =  Luminosity $\times$ Duration, it follows 
that the duration is essentially independent of the  
luminosity of the event.  
As far as can be determined from this sample moreover,
the energy and luminosity of flares are independent of age.

\subsubsection {Frequency of Flares}

We found 41 large flares during an average effective
exposure time of 660 ks on 27 observed solar mass stars 
(COUP\,828 was excluded from flare searches because of its proximity
to a chip gap).
Thus, we compute a preliminary average of 1 flare per star per $\sim$ 
435 ks of observing time. 
The true rate of flares is somewhat lower than this; however, because
the effective exposure time on some sources was
limited due to dither and off--axis effects.  While such effects lower
the total number of counts observed, the overall sensitivity to flares 
is not compromised because all flares observed exceed the duration of the 
dither cycle.  We argue that we are sensitive to flares that
occurred on any of the 27 stars during the full 850ks observation.
Further, flares 
which occurred when the ONC was not being observed  
would still have been detected
as such if the flare lasted into the next observing window. 
Although we cannot detect any flares that take place entirely during  
perigee passage by $Chandra$,  we do detect flares that begin
during the perigee passage and extend to about 8 hours after perigee,
when new observations start.
Since the median flare duration is about the same length
as perigee passage, we miss about half of the flares that begin during
perigee.
The time between observations is about 57~ks, and there were five
perigee passages during the observations, so we were sensitive to
one half of the flares occurring during about 288 kiloseconds
of perigee passages. 
Applying these corrections, we infer an average of 
1 flare per star per 650 ks.  This is consistent with earlier results for Orion 
and much older clusters as reported by Wolk \e (2004).  However
the flare definitions used were less
quantitative than those used here and biases in the data are
unclear.  Specifically, Wolk \e (2004) were more sensitive to
faint flares that would be considered elevated in the current
work.  Nonetheless, it appears that flare rates change by a factor
less than 5 during the first 100 Myrs.

Two stars, COUP\,1023 and 1281, show no evidence of flaring.  
COUP\,241, 1167 and 1159 only showed flaring in the extremely early or late
observation intervals. Eleven stars underwent two flares,
one star was observed to have three,  and another exhibited four.
{\it A priori}  
we did not consider the quiet nature of  COUP\, 1023 and 1281,  or the active
nature of  COUP\, 314 and 567, anything more than statistical fluctuation in a
random distribution of flares.
On the Sun however, the observed power-law distribution of flare energy
release is well characterized by the
assumption that the solar corona is critically
self-organized (Lu \& Hamilton 1991).  A result of this organization is
a waiting time distribution (Lu \e 1993) between flares that is
power law in nature with a slope of  $f(t) \propto  t^{\Gamma ~t}$ where
the index $\Gamma$ is about 2.15-2.55 (Wheatland 2000, Norman \e 2001).

To test whether stars in the sample were
subject to a nonrandom waiting time, 
we assumed that all stars in the sample have 
the same behavior (and thus that observing 27 of them for 850 ks is 
equivalent to observing one  of them for 24 Ms -- the ergodic 
theorem).
The distribution was simulated using a simple  
Monte Carlo model for the distribution of
flares among the 27 stars.  For each star, the lightcurve was divided 
into 39 25~ks bins, assigned a random number to each bin, if
the random number exceeded a threshold, we noted this as a flare.  We set the
threshold to trigger once per 625 ks (25 bins) on average.  One
thousand datasets of 27 stars were simulated, each 
with about 41 flares in each
dataset.  The results of the simulation are shown in Figure~\ref{monte}.
Not surprisingly, the result is a Poisson distribution centered between 1 and 2
flares per star with a few extra sources in the zero flare bin 
since it is impossible to have negative flares.  
A two-sided KS test fails to discriminate the 
ergodic model from other models 
(probability that the data fit the ergodic hypothesis was 44\%).
Thus, at the level of our data, there is no indication that the
temporal distribution of flaring is anything other than random.

\subsubsection{Shape of Flares}
\skipthis{
The shape of the flares may indicate whether a flare is associated
with a coronal mass ejection, the isotopic ratios within the CME, and thus
whether the flare can trigger various types of {\it in situ} nucleosynthesis.
In observations of solar flares ($E\ga 1$ MeV per nucleon),
there are two distinct phenomena associated with solar energetic
particles (SEP) events: ``impulsive'' SEP events, and ``gradual'' SEP
events. 
The impulsive SEP events are characterized by
their unusual abundances relative to coronal
abundances. 
Enhancement factors of $\sim 1000$ in $^3He/^4He$ (Reames et al 1999)
are observed in the Sun.
The gradual SEP events do not carry unusual elemental abundances
in the composition, and on average the compositions are similar to
those of the corona and solar wind.}

In our initial qualitative analysis of the bright flares in the
dataset, we categorized flares by luminosity, duration and shape.
Specifically, the shape was noted as either, ``linear rise plus
exponential decay'', ``symmetric'' or ``spike'' (i.e., less than 5\,ks from
beginning to end). There were also a handful of flares which were too
weak to describe.   Eight flares were noted as
symmetric.  The first flare of COUP\,1259 is probably  the best
example of a slow-rise roughly symmetrical flare.
In two sources, COUP\,131 and 1570,
    multiple symmetric events may be indicative of rotational
    modulation of long-lived structures in the stellar corona, rather
    than individual magnetic reconnection events (these are detailed
by Flaccomio \e 2005).
    Three of the remaining five symmetric events are too short-lived
    to be rotational events (durations $<$1 d) and are not seen to repeat.

\subsection{Temperature} 

The temperature profile of the flares is critical to understanding
their ability to penetrate circumstellar material.  Harder
X-rays penetrate more material and can heat deeper into disks'
midplanes.  A fundamental question is whether or not the flare plasmas 
of these T Tauri stars share properties known in flare plasmas 
of older stars, including:     L$_x \propto$ kT$_1^{2.2}$
(Preibisch 1997), 
 L$_x \propto$ kT$_2^5$ (G\"uedel et al. 1997) 
and  kT$_1$ is always around 0.8 keV (Sanz-Forcada \e 2003)  
where T$_1$ and T$_2$ are the cooler and hotter 
plasma components, respectively.

We have fit most of the X-ray sources as two temperature MEKAL models.
We find a significant correlation between the hot and cool components
of the characteristic X-ray emission as shown in Figure~\ref{kt1kt2}.  
A similar effect was seen in ROSAT data albeit with a different
instrument and for different stars
(cf. Stern \e 1994, Gagn\'{e}  ~\e 1995 and Figure~24 of Favata \& Micela
2003). The effect has also been noted in the time-integrated spectra of COUP
stars (Preibisch \e 2005a).   

At the characteristic levels, the soft component is weakly correlated
with the characteristic luminosity  (correlation coefficients
indicate $< 0.15$ probability of no correlation) and the
harder component shows no correlation whatsoever ($\tau;\rho; \sim 0.90$).
However, the characteristic hot and cool coronal components  
are highly correlated: a linear regression fit gives
$kT_2 = 2.14 \times kT_1 + 0.66$ keV, with a median deviation of 0.5keV
about the fit.
A key difference between this data set and the ROSAT results mentioned above is
the hard sensitivity and corresponding lack of soft X-ray sensitivity 
of the Chandra ACIS-I.  Thus, it is not surprising that we find a hard
component  median of 
about 2.3~keV and a median soft component around 0.9~keV, both much
warmer than found with ROSAT.

The temperature correlation breaks down during the elevated and flare
periods ($\tau,\rho \sim 0.25$ for both flare and elevated periods). 
It is clear that the
temperature of the hot component of the plasma increases  
as a star's flux level changes from the characteristic level
to the flaring state; the hot plasma achieves maximum heating when the star is at
peak flux levels.  The cool components do not change significantly. 
This evidence suggests that the breakdown in
the correlation occurs because the flare is only manifest in the hot corona.
The median of the hot component increases by over 50\% to 3.45 keV
during flares, and five flares
exceeded temperatures of 100~MK (9~keV) in their hot component.
Favata \e (2005) examined the brightest flares in the COUP
sample and found 100MK temperatures in half of the events.  This is
supporting evidence for a L$_x$--kT$_2$ relationship. Further, it
implies that hotter flares than those witnessed during this 13.2 day observation
probably occur on these stars.
The cool component shift (formally from 670 eV to 760 eV) is consistent with a
constant kT$_1$ as found by by Sanz-Forcada \e (2003).

\subsection{Stars With and Without Disks}
We have reasonable disk and flare data on 26 of the 28 stars in the
sample but lack $\Delta (I-K)$ measurements for two
stars (Table~2).
Three stars have evidence of accreting disks via \ion{Ca}{2} emission lines, 
while 13 have evidence of passive accretion disks as ascertained by a K band excess or
direct proplyd detection without  \ion{Ca}{2} emission lines.  Ten stars 
have no evidence of a disk.
Among the active accretors, two are faint with a single weak flare,
and one, COUP\,567, has multiple strong flares. The 
group of stars with passive accretion disks include the 
two obvious rotators: the star with four flares and another star with
no flares.
The subgroup without evidence of any disk also contained one star
without flares and its remaining stars had one or two modest flares. 
There is thus no evidence dependence of flare
  properties on the presence of dusty disks or accretion.  This issue
  will be revisited in a future COUP paper using larger samples.

\section{X-ray Interactions with Circumstellar Disks}
	As is the case for all YSOs,  X-rays from the Sun-like stars
in the ONC affect
conditions in their immediate environment (see e.g., the review by GFM00). 
In a dense cluster, nearby stars may contribute, but close to any given
star, the X-ray emission from that star (and any companion)
will dominate. Having found that solar mass stars have a
characteristic X-ray luminosity, it is appropriate to update the formula
given by GFM00 for the ionization rate $\zeta$ at a radial distance $r$ (in
units of AU) from a YSO (see also Glassgold, Najita, \& Igea 2004),
\begin{equation}
\zeta = {6\times 10^{-9}} ~~\ps \, (\frac{2 \times 10^{30}}
{L_{char} ~[\erg\,\ps]})\, (\frac{\rm AU}{r})^2.
\end{equation}
This formula ignores likely attenuation and scattering of the X-rays,
and uses the 
median characteristic X-ray luminosity given in \S 6.1, 
which is appropriate for solar mass stars; this value is lower for more
common lower mass stars.

In their calculation of the production of short-lived radionuclides in
the region near and inside the inner disk or co-rotation radius, Lee
et al. (1998) invoked observations of soft and hard YSO X-rays made
with ROSAT and ASCA to estimate the fluence of nuclear particles. They
converted X-ray to stellar energetic-particle fluxes using
observations of the contemporary active Sun. A related calculation by
FGP02  
using $Chandra$
observations estimated that the particle fluxes from active
YSOs were $\sim 10^5$ more intense than in the active Sun. This
number reflects the fact that YSO flares are more powerful and more
frequent than those of the contemporary Sun, and that the energy 
distribution of solar energetic particles is shallower than that for
X-rays. FGP02 found that the inferred increase of $10^5$ in particle
fluence is more than sufficient to explain by spallation 
the abundances of several important meteoritic isotopic anomalies 
(Woolum \& Hohenberg 1993, 
Lee et al. 1998, Goswami et al. 2001, Gounelle et al. 2001, 
Marhas et al. 2002, Leya et al. 2003).

	Our COUP observations of solar-mass YSOs can be used
to make similar estimates of the particle fluence in the reconnection
ring. If we take the mean characteristic luminosity as
$2 \times 10^{30}$\,\erg\,s$^{-1}$, and the mean flare luminosity, duration, 
and repetition times 
as $6 \times 10^{30}$\,\erg\,s$^{-1}$, $10^5$\,s, and 650~ks
respectively, the fluence  at a distance $0.75 R_x$ 
(with  $R_x = 0.05$\,AU, the x-point or co-rotation radius) over 10
years becomes: 
\begin{equation}
{\cal F}_{\rm X}(10\,{\rm yr}) = 2 \times 10^{15}  ~~\erg\, \psqcm.
\end{equation}
If we now convert from X-ray to proton fluence (for energies 
greater than 10 MeV) using the same 0.1 factor as used by Lee et al. (1998), 
we get essentially the identical result as those authors,
\begin{equation}
{\cal F}_p(10\,{\rm yr}) = 2 \times 10^{14} ~~\erg\, \psqcm.
\end{equation}
Individual stars will exhibit a scatter of a factor of $\pm 3$
  or more about these values.
A caveat to be discussed further in the next
paragraphs is that the nuclear irradiation of calcium-aluminum inclusions
(CAIs) probably occurred at an earlier
 stage of pre-main sequence evolution than seen in the
  unobscured ONC stars considered here.

 A potentially important connection permitted by 
the COUP's 13.2--day
exposure is that very short ($\lesssim 1$\,hr) as well as long 
duration ($\gtrsim 1$\,d) flares observed are reminiscent of solar 
impulsive and gradual flares. 
Although the COUP data cannot directly detect the brief
    very-hard spectral signature of impulsive phases at the
    beginning of ONC flares, we can infer by
    analogy with the relationship of solar impulsive and gradual
    flares that the impulsive phase is often present.  
This may be
very significant since, 
on  the contemporary Sun, impulsive and gradual phases of solar flares
flares have different properties, most strikingly in their elemental
and isotopic composition. Unlike proton-rich gradual events, which are
also associated with coronal mass ejections, impulsive events are
$^3$He rich, and are therefore particularly effective in producing
some short-lived radionuclides like $^{26}$Al and $^{41}$Ca, but not
 $^{60}$Fe. (Lee \e 1998, Gounelle \e 2001).

	As basic as the nuclear fluence estimates are, the YSO X-rays can
also directly affect the physical state of the irradiated material,
i.e., the proto-CAIs and chondrules seen
in the earliest solar system solids.
Favata \e (2005) examined two of the stars in our sample COUP\,223
and COUP\,262 and fit these events to coronal loop models.   
They find that the flare in COUP\,262 extended at
least 3.6R$_*$ and may have reached 18R$_*$ indicating possible direct
contact between the inner portion of the disk and magnetic loop of
the flare.
According to Lee et al. (1998) and Gounelle et al. (2002), solids
experienced thermal processing episodes for several years before being
launched into the primitive solar nebula. These events induced a
variety of phase changes, including partial or full
evaporation. Shock waves from these powerful flares may also
propagate along the outer layers of the protoplanetary disk and melt
proto-chondrules several A.U. from the star (Nakamoto et al. 2005).
Powerful flares, such as those at the high end of the
distributions in Figures \ref{lum_hist}, \ref{ratio} and \ref{energy},
 may thus have important effects on the solar nebula
disk material in several ways.  Their hard penetrating X-rays will
ionize disk gas, even into the midplane.  Their energetic baryonic
particles may produce short-lived radionuclides in disk solids via
spallation with their energetic particles.  Their thermal flashes or
shock waves may melt disk dustballs into CAIs or
chondrules.

This particular application also points out an important limitation of the
$Chandra$ observations that will be difficult to overcome without a
significant enhancement in the capability of X-ray observations. Most
of the YSOs observed in the ONC, including the solar-analog sample of
this paper, are revealed T Tauri stars with a median age of the order
of 2~Myr. Although a fair fraction have disks and are still
accreting, most of the stellar mass has already been accreted,
following an earlier more active stage of accretion that probably
occurred during the first several hundred thousand years of their
lives. It is very likely that the nuclear irradiation that led to some,
if not many, of the short-lived radionuclides occurred during this
early period and not in the T Tauri phase. 
We note that the meteoritic evidence for high energy processes
that we attribute to X-ray flares takes place over an extended and complex
period.  CAIs are irradiated and melted during a brief ($<10^4$ yr)
protostellar phase, chondrules are melted over a somewhat longer
phase ($10^5$ yr), while grains showing spallogenic $^{21}$Ne
excesses are irradiated over $10^7$ years.  Although the sample of
young solar analogs studied here was selected to be unobscured $\sim
10^6$ year old stars, we note that both the younger COUP stars
embedded in the OMC-1 cloud cores (Grosso et al. 2005) and older
COUP stars ($\simeq 10^7$ yr; Preibisch \& Feigelson
2005) exhibit very similar X-ray luminosity functions and flaring
behavior as the sample examined here.  It is perhaps uncertain only
whether or not X-ray emission extends to the very youngest Class 0
protostars.
\skipthis{
However, COUP has revealed many of the stars around
    the two OMC-1 molecular cloud cores -- the
    Becklin-Neugebauer-Kleinman-Low and OMC-1 South regions -- which
    are younger than ONC stars and likely in their protostellar phases
    (Grosso \e 2005).  These very young, deeply embedded stars show the
    same X-ray luminosity function and same flaring properties as the
    optically revealed ONC stars. This indicates that the X-ray
    properties discussed here for ONC solar analogs applies during
    earlier phases, though it is unclear whether it applies to the very
    youngest Class 0 protostars.}

\section{Summary and Conclusions}

We have examined a sample of 28 solar--mass stars observed by 
$Chandra$  during a very deep observation and have  
quantified the lightcurves of the X-ray sources using maximum 
likelihood analysis.  Taking advantage of the
extremely long 
and nearly continuous nature of the observation we applied quantifiable  
definitions to the lightcurves to identify characteristic levels,
flares and intermediate states for 27 of these sources.
Principle among our findings is that young solar analogs spend about three-fourths of
their time in a relatively well behaved characteristic state.  This
state can be well modeled as a 2--temperature plasma with one
component fixed near 850 eV and the other component of about 2.35 keV.
This characteristic state is marked by good
correlations among their coronal temperatures and bolometric to X-ray
luminosities, but these relationships break down in the more active states.  
While the flares are well
fit by $dN/dE \propto E^{-1.7}$, a power law index of $\sim$ -2.25 is required to 
match the shape of the overall lightcurves if they are purely the result
of flaring.  This broken power law is
consistent with recent observation of RS CVn stars (for bright flares)
and AD Leo  (for fainter flares) and indicates that the characteristic
portion of the luminosity does not arise from flares (or at least not
the same kind of flares that give rise to the luminosity in the
flaring state). 
The emission of the stable component represents about 0.03\% of the
bolometric luminosity.

Some of our conclusions regarding behavior during the very complex 
flaring state are as follows:
\begin{enumerate}

\item The median flare has a luminosity of about 10$^{30.8} \erg\,\ps$
and releases about 10$^{35.5}$ ergs with kT$_{hot}\sim
3.45~$keV.  
The peak luminosity of flares is almost 10$^{31} \erg\,\ps$
and lasts about 10ks.  During the peak of the flare kT$_{hot}$ has a
median of about 7~keV. 

\item The ratio of the flare peak to characteristic flux is 
less than a factor of 100. Flares induce a large change 
in the temperature of the hot coronal plasma component. 

\item We find no relation between luminosity and duration, nor 
do we see significant trends of luminosity with age.

\item There is no statistical evidence that flares are subject to 
a waiting time between flares.  The time between events can be 
modeled as a random occurrence.  The duration of the flares themselves 
varied from less than an hour to almost three days.

\item The cool component of the corona is unaffected by activity, with
the median value of kT$_{cool}$ remaining between 710 and 900 eV in all states.
\end{enumerate}

The intensity and hard spectra found for ONC solar
    analogs indicate that the resulting ionization of disk gases by
    stellar X-rays dominates ionization by cosmic
    rays or other sources by a large factor. 
        If solar-type particle production is associated with thermal
    flares, the observed flaring rate inferred for the early Sun is
    sufficient to produce many of the isotopic anomalies seen in
    meteoritic inclusions.

X-rays do not account for all anomalies seen in meteoritic inclusions or 
protosolar analogs.   Although X-ray heating of gas above the
midplane is significant and  rotation--vibration transition of CO should
be easily excited.  But this does not occur at rates significant
enough to explain recent observations of very strong CO signature in
some T Tauri stars.

\acknowledgments
We thank the referee Alex Brown for a critical reading and helpful
suggestions regarding all aspects of the manuscript.  We are extremely
grateful to A. Glassgold for helping to lay out the arguments in the
second to last section and several critical readings.
We thank Salvatore Sciortino (Palermo) for a
critical reading of this paper and Jay Bookbinder (SAO) for a
valuable discussion.  This research made use of SAOImage DS9,
developed by the Smithsonian Astrophysical Observatory. This research
also made use of the Two Micron All Sky Survey (2MASS), a joint
project of the University of Massachusetts and the Infrared Processing
and Analysis Center/California Institute of Technology, funded by the
National Aeronautics and Space Administration and the National 
Science Foundation. COUP is supported by Chandra Guest Observer
grant GO3-4009A (E.\ Feigelson, PI). S.J.W. and F.R.H. received
support from $Chandra$ X-ray Center contract NAS8-39073 and HRC
contract NAS8-38248, E.D.F. received support from $Chandra$ ACIS
contract NAS8-38252, E.F, G.M. and S.S. received support from the
{\em Ministero dell'Istruzione dell'Universit\'a e della Ricerca}.

Facility: CXO(ACIS) 
\dataset[ADS/Sa.CXO#obs/COUP] 
\object[ADS/Sa.CXO#obs/COUP]{17},
\object[ADS/Sa.CXO#obs/COUP]{54},
\object[ADS/Sa.CXO#obs/COUP]{57},
\object[ADS/Sa.CXO#obs/COUP]{131},
\object[ADS/Sa.CXO#obs/COUP]{147},
\object[ADS/Sa.CXO#obs/COUP]{177},
\object[ADS/Sa.CXO#obs/COUP]{223},
\object[ADS/Sa.CXO#obs/COUP]{241},
\object[ADS/Sa.CXO#obs/COUP]{250},
\object[ADS/Sa.CXO#obs/COUP]{262},
\object[ADS/Sa.CXO#obs/COUP]{314},
\object[ADS/Sa.CXO#obs/COUP]{515},
\object[ADS/Sa.CXO#obs/COUP]{567},
\object[ADS/Sa.CXO#obs/COUP]{753},
\object[ADS/Sa.CXO#obs/COUP]{828},
\object[ADS/Sa.CXO#obs/COUP]{1023},
\object[ADS/Sa.CXO#obs/COUP]{1127},
\object[ADS/Sa.CXO#obs/COUP]{1134},
\object[ADS/Sa.CXO#obs/COUP]{1151},
\object[ADS/Sa.CXO#obs/COUP]{1167},
\object[ADS/Sa.CXO#obs/COUP]{1235},
\object[ADS/Sa.CXO#obs/COUP]{1259},
\object[ADS/Sa.CXO#obs/COUP]{1281},
\object[ADS/Sa.CXO#obs/COUP]{1326},
\object[ADS/Sa.CXO#obs/COUP]{1327},
\object[ADS/Sa.CXO#obs/COUP]{1500},
\object[ADS/Sa.CXO#obs/COUP]{1539},
\object[ADS/Sa.CXO#obs/COUP]{1570}

\clearpage

\begin{figure}[tbph]
\vspace{3.0in}
\centerline{\Large\bf Figure 1 is not available in the astro-ph version}
\vspace{3.0in}
\caption{The locations of the Solar mass stars in Orion.
A- The Coup image of the ONC 17\arcmin~ on a side 
(Blue = 2.8-8 keV, Green 1.7-2.8 keV, Red= 0.5-1.7 keV).
Locations of Solar mass stars are circled.  The two solar analogs not
detected are noted with ``JW''. 
B -  The 2MASS image of the central 12\arcmin~ of the ONC
(Blue = J band, Green = H band, Red= K$_s$ band).
Locations of Solar mass stars are circled.\label{image}}
\end{figure}


\begin{figure}[t]
\plotfiddle{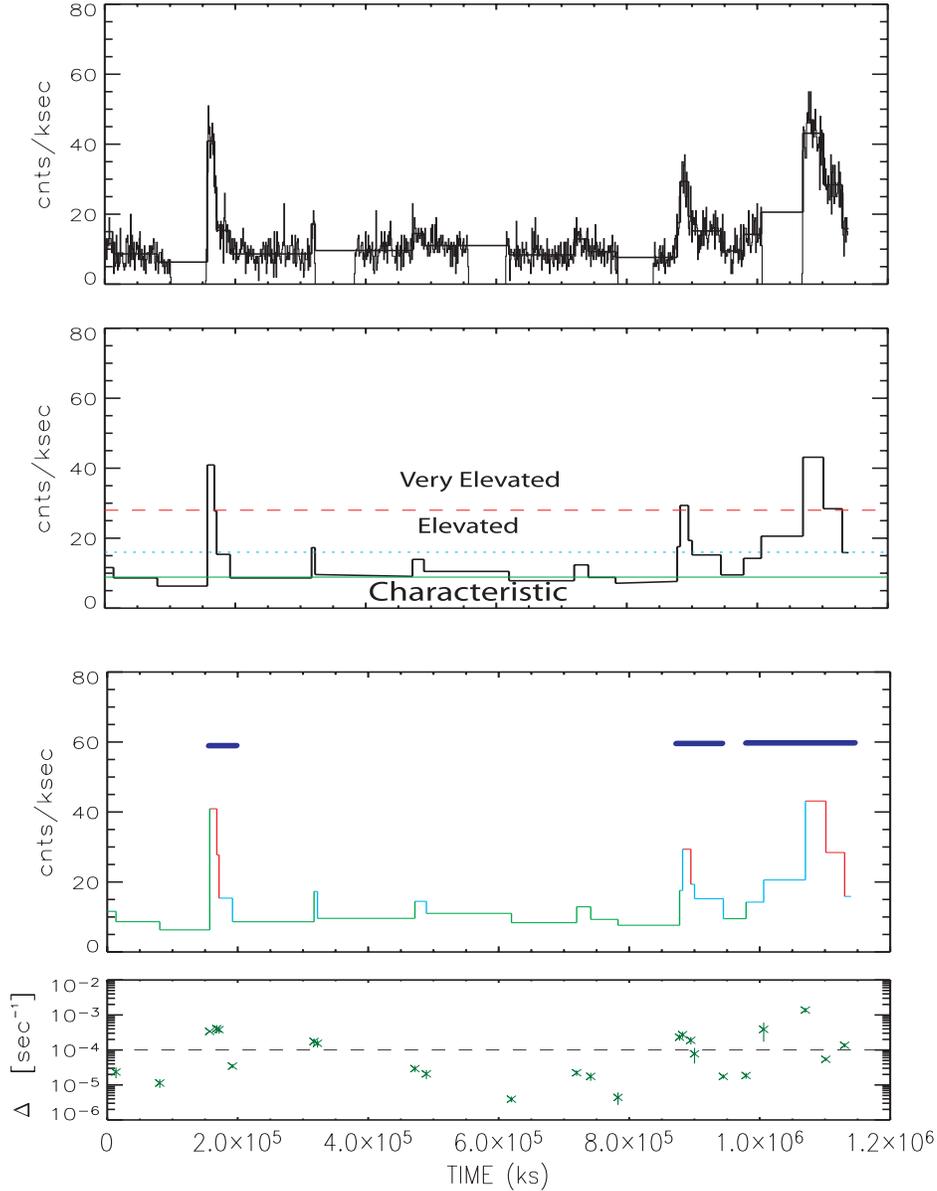}{-1.0in}{0.}{350}{450.}{50.}{0}
\caption{Definition of variability levels in a lightcurve of a typical 
solar-mass COUP source.   Top:  the lightcurve,
represented by a histogram with 1 hour time bins, is converted to
blocks of constant flux.   Second panel: The characteristic
level is determined as shown by the green solid line.  The elevated
level 
is marked by the dotted cyan line. The very elevated level
is marked by the red dashed line. 
Every block below the elevated line is grouped with the characteristic
blocks.  Every block above the very elevated line is in the very
elevated group, the remaining blocks are considered elevated.
Third Panel:  Blocks are colored as in the second panel.  Flares as 
determined in the bottom panel  are indicated by the blue lines.
Bottom: The rate of change $1/R_{char}\times dR/dt$ is calculated for
each block interface, for this case, all three excursions into the
very elevated regime are accompanied by rapid changes and are thus
considered flares.  Note that some rapid changes are not accompanied by
high flux rates and thus are not considered to be flares.}
\label{sample}
\end{figure}
\clearpage

\begin{figure}[t]
\vspace{3.0in}
\centerline{\Large\bf Figure 3 is not available in the astro-ph version}
\vspace{3.0in}
\caption{COUP lightcurves for the 28 solar-mass Orion Nebula Cluster stars.
COUP source numbers appear above the plots.  The top panel shows 
the $0.5-8$ keV brightness variations with bin sizes ranging from 1 ks
to 10 ks.  Binning is for display only was not used in the analysis.  
Horizontal lines show the MLB segments.  In the electronic version, 
these are colored green, cyan and red for characteristic, elevated 
and very elevated segments, respectively.  The horizontal blue lines 
above some red regions indicates periods of flare levels.  The lower 
plot in each panel shows the rate of change between adjacent segments 
with a dashed line indicating the criterion for flares.  The abscissa
of  all plots show time in seconds from the beginning of the COUP 
observation.  The observation is 13.2 days long with five gaps due to 
$Chandra$ orbit perigees. \label{lc14}}
\end{figure}
\clearpage

\begin{figure}[tbph]
\plotone{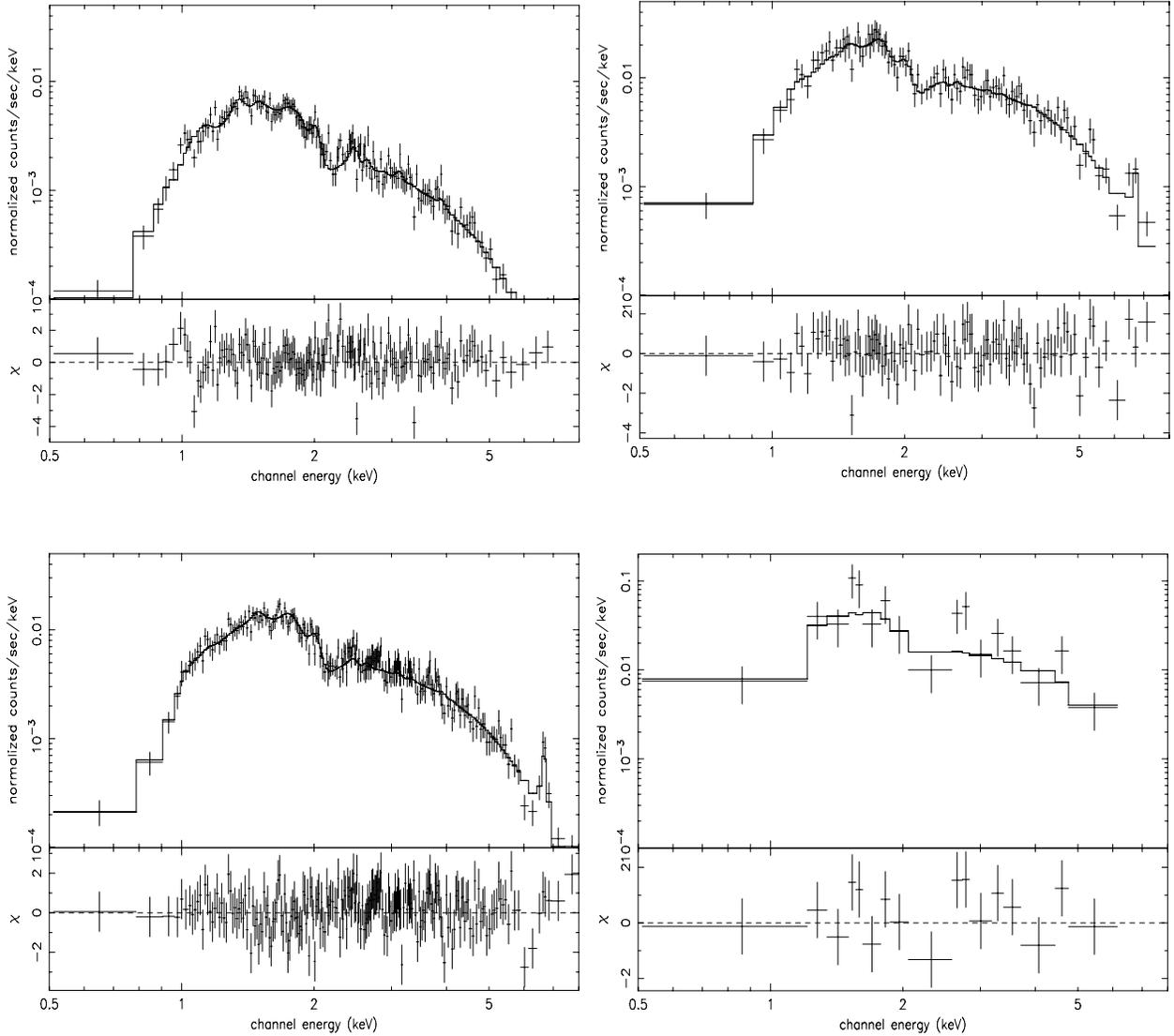}
\caption{An example of fitted spectra of COUP \# 262 with \nh  
fixed to 1.6 $\times 10^{21} $cm$^{-2}$ (in the text it is a free parameter).
Upper left: photons which arrived during the characteristic period
(kT$_1$ = 3.49eV, kT$_2$ = 3.51keV).
Upper right -- photons which arrived during the elevated period 
(kT$_1$ = 80eV, kT$_2$ = 5.96keV).
Lower left -- photons which arrived during the flare
(kT$_1$ = 80 eV, kT$_2$ = 8.64keV).
Lower right -- photons which arrived during the peak flux block of the
flare (kT = 27.1keV).
All fits are two temperature MEKAL fits except for the lower right.
\label{sampleq}}
\end{figure}

\begin{figure}[tbph]
\plotfiddle{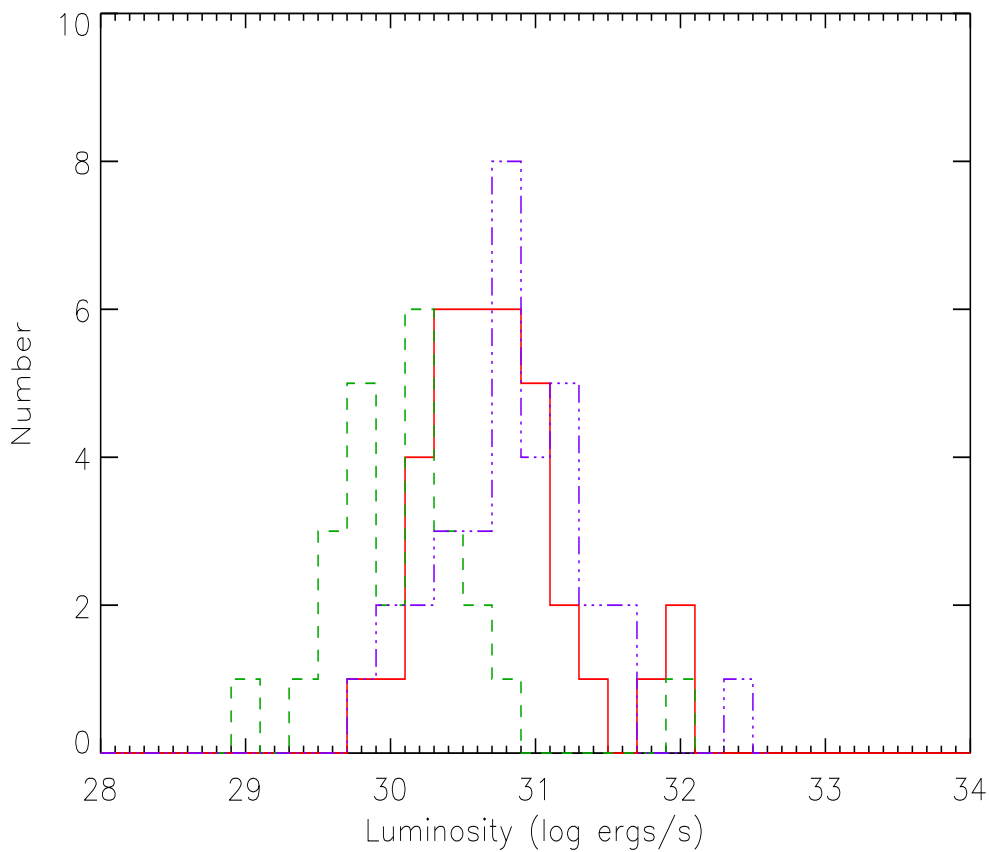}{-2.0in}{90.}{400.}{470.}{0.}{00.}
\caption{Histograms of the characteristic (dashed line),
composite flare (solid line) and peak flare  (dash-dotted line)
luminosities of Solar mass stars in the ONC; Only data with 
\nh fixed to 1.6 $A_v \times 10^{21} ~cm^{-2}$ are plotted. The characteristic
luminosities are relatively evenly distributed across two orders of magnitude,
centered near $\log L_{char} \sim 30$. The flare
luminosities are more sharply peaked near $\log L_{flare} \sim 30.5$.
 The peak flare luminosities are centered near  $\log L_{flare} \sim 31$.
\label{lum_hist}}
\end{figure}

 \begin{figure}[tbph]
\plotfiddle{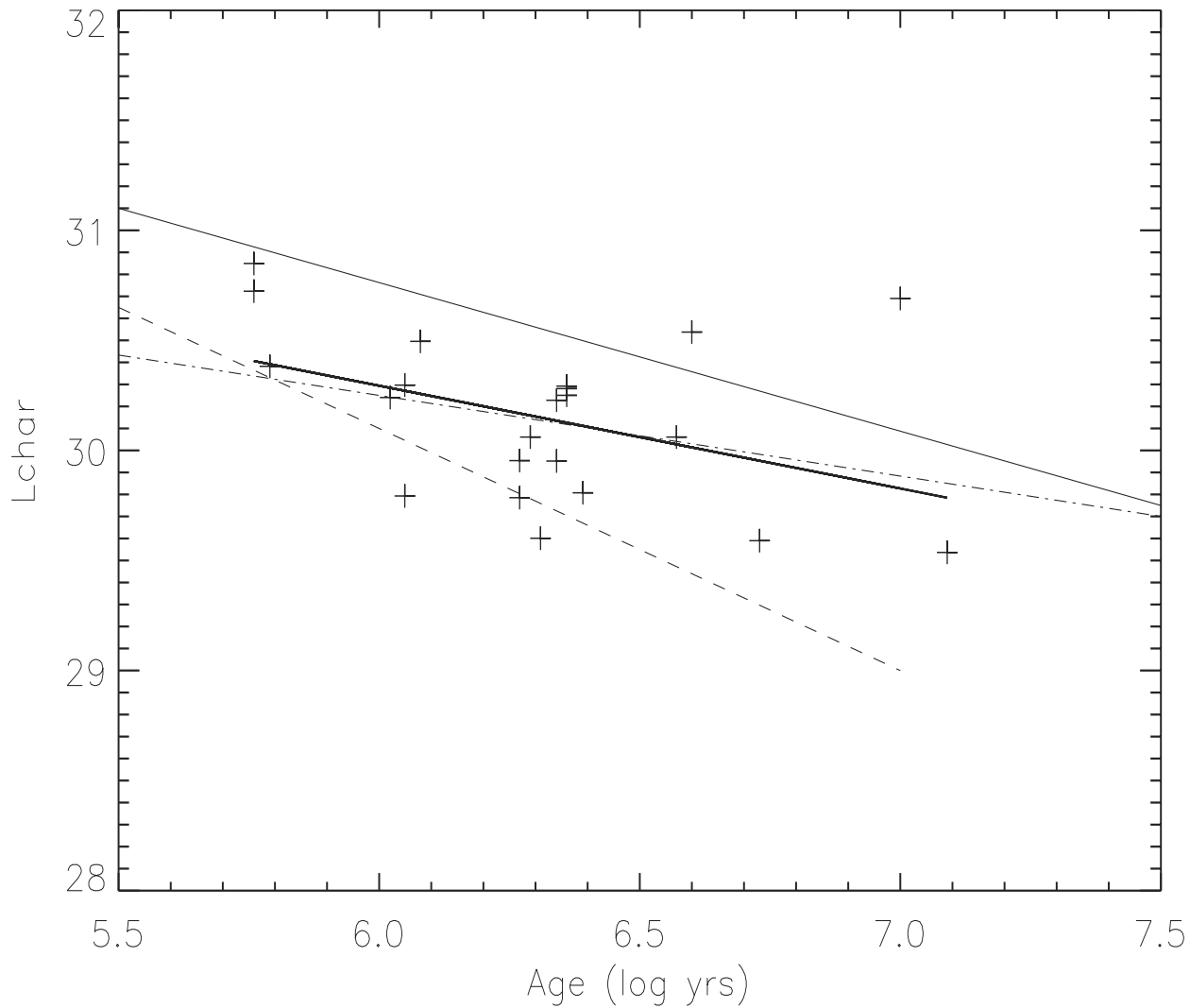}{0.0in}{90.}{400.}{470.}{0.}{-500.}
\caption{Characteristic X-ray luminosity of Solar mass stars in
Orion as a function of age.  
The dashed trend line is adapted from FGP02 and the thick solid trend line is
an outlier-resistant  two-variable linear regression
to the data presented here.  The dot-dashed and thin solid lines are taken from
 Preibisch \& Feigelson (2005) Figure~1 for mass ranges of 0.4--1M$_\odot$ and
1--2M$_\odot$.  The  0.4--1M$_\odot$ mass range matches the fit derived here.
\label{lt_age}}
\end{figure}

\begin{figure}[htp]
\plotfiddle{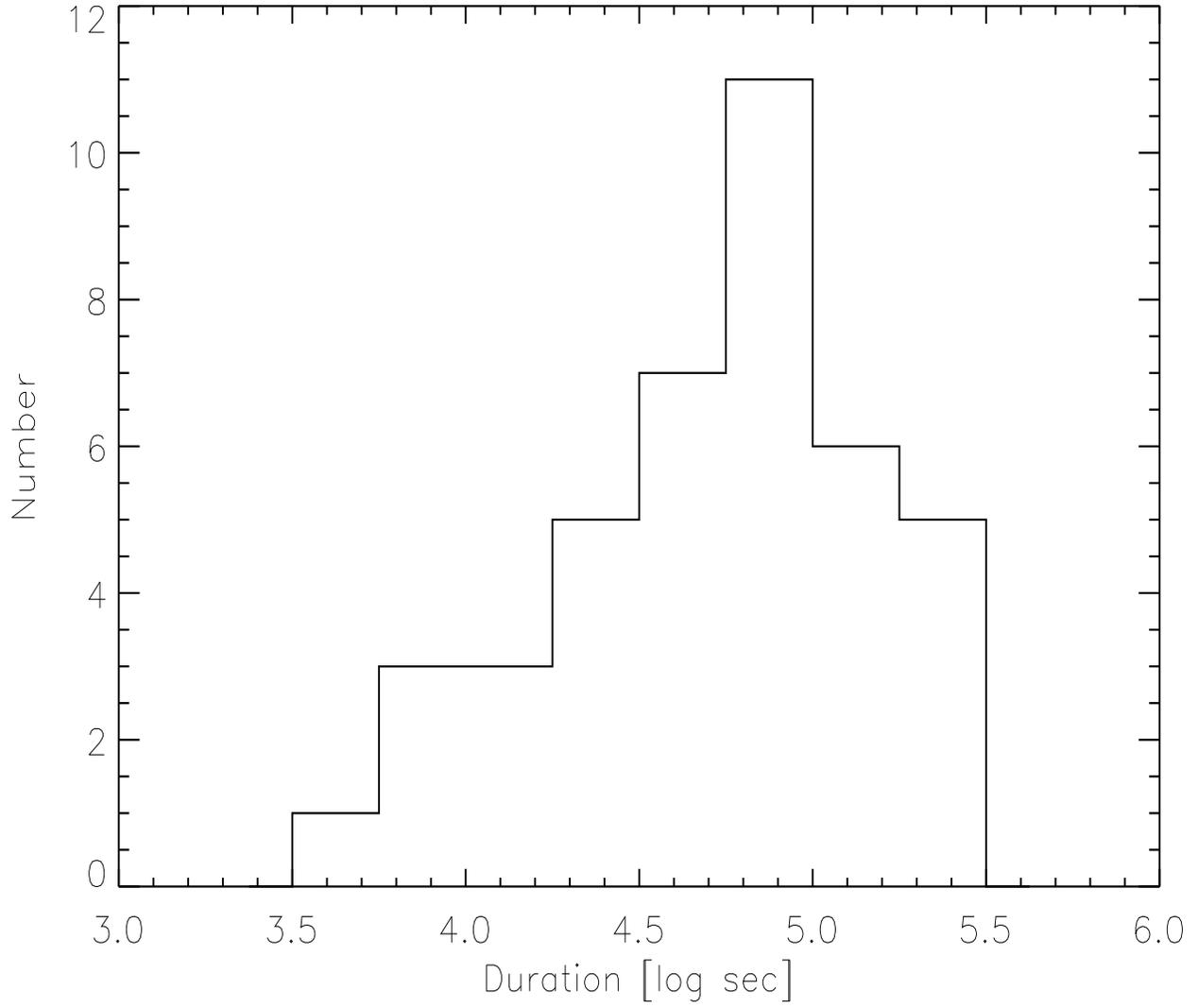}{-4.0in}{90.}{400.}{470.}{0.}{00}
\caption{Distribution of the duration the ``whole flare,'' (see text)
for all 41 flares.
 \label{duration}}
\end{figure}

\begin{figure}[tbph]
\plotfiddle{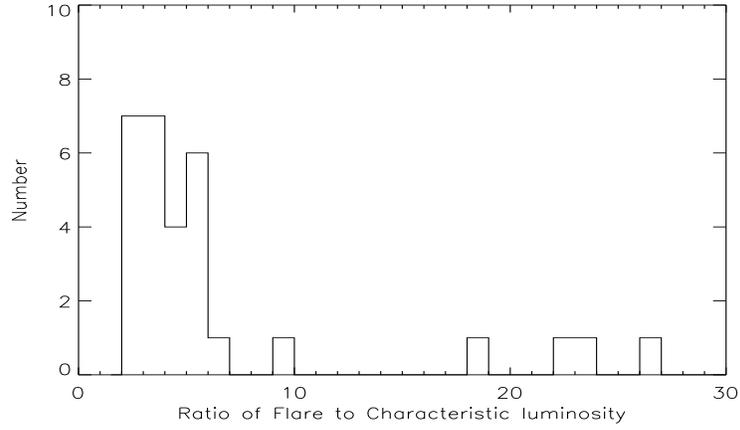}{0.0in}{90.}{200.}{350.}{50.}{500.}
\plotfiddle{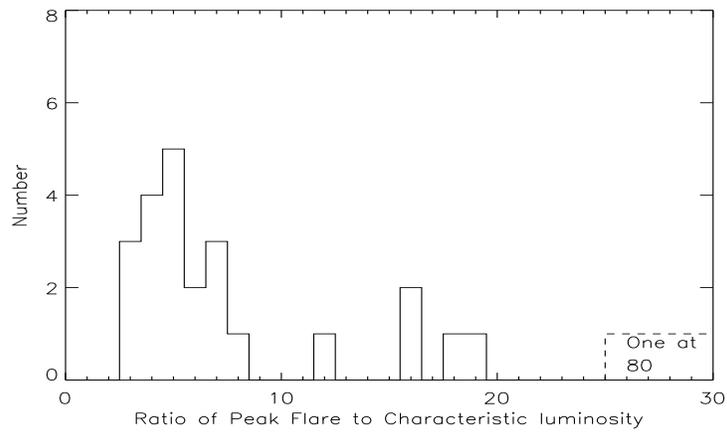}{0.0in}{90.}{200.}{350.}{50.}{500.}
\caption{Strength of flares.  
Top: Ratio of the {\bf average} luminosity
of  each flare to the  characteristic level for that star, using the
``whole flare'' criterion (see text).
Bottom: Ratio of the {\bf peak} luminosity for
each flare to the characteristic level for that star.
Fewer sources are plotted in the shown plot because the ``peak flare''
criterion sometimes yields insufficient counts for a reliable fit.
\label{ratio}}
\end{figure}

\begin{figure}[htp]
\plotfiddle{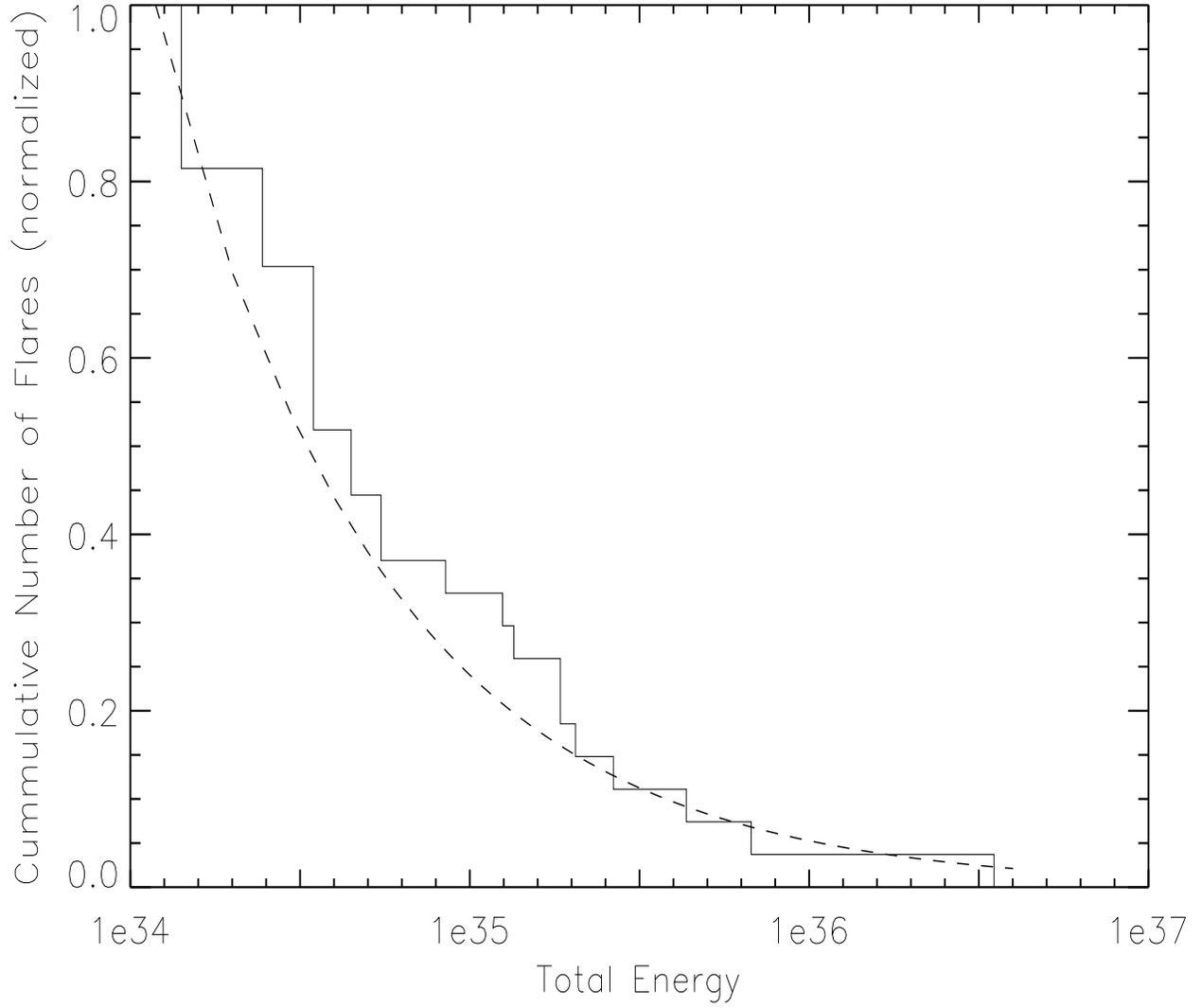}{-4.0in}{90.}{400.}{470.}{0.}{00}
\caption{Cumulative distribution of flare energy for 27 flares with
good spectral fits.  The dashed line is the best fit curve
N=1.1 $\log E^{-0.66}$.
 \label{energy}}
\end{figure}

\begin{figure}[htp]
\vspace{3.0in}
\centerline{\Large\bf Figure 10 is not available in the astro-ph version}
\vspace{3.0in}
\caption{Simulated lightcurves for various power law
spectra for the count rate of flares. Two values for the
total number of flares are shown.
The figures show the resultant lightcurves.
The short (red) lines indicate the individual flares. 
Overall, the simulated lightcurves are
remarkably similar to the actual observed light curves. Power laws
between -2 and -2.5 seem most similar to the observed data.
\label{sims}}
\end{figure}

\begin{figure}[tbph]
\plotfiddle{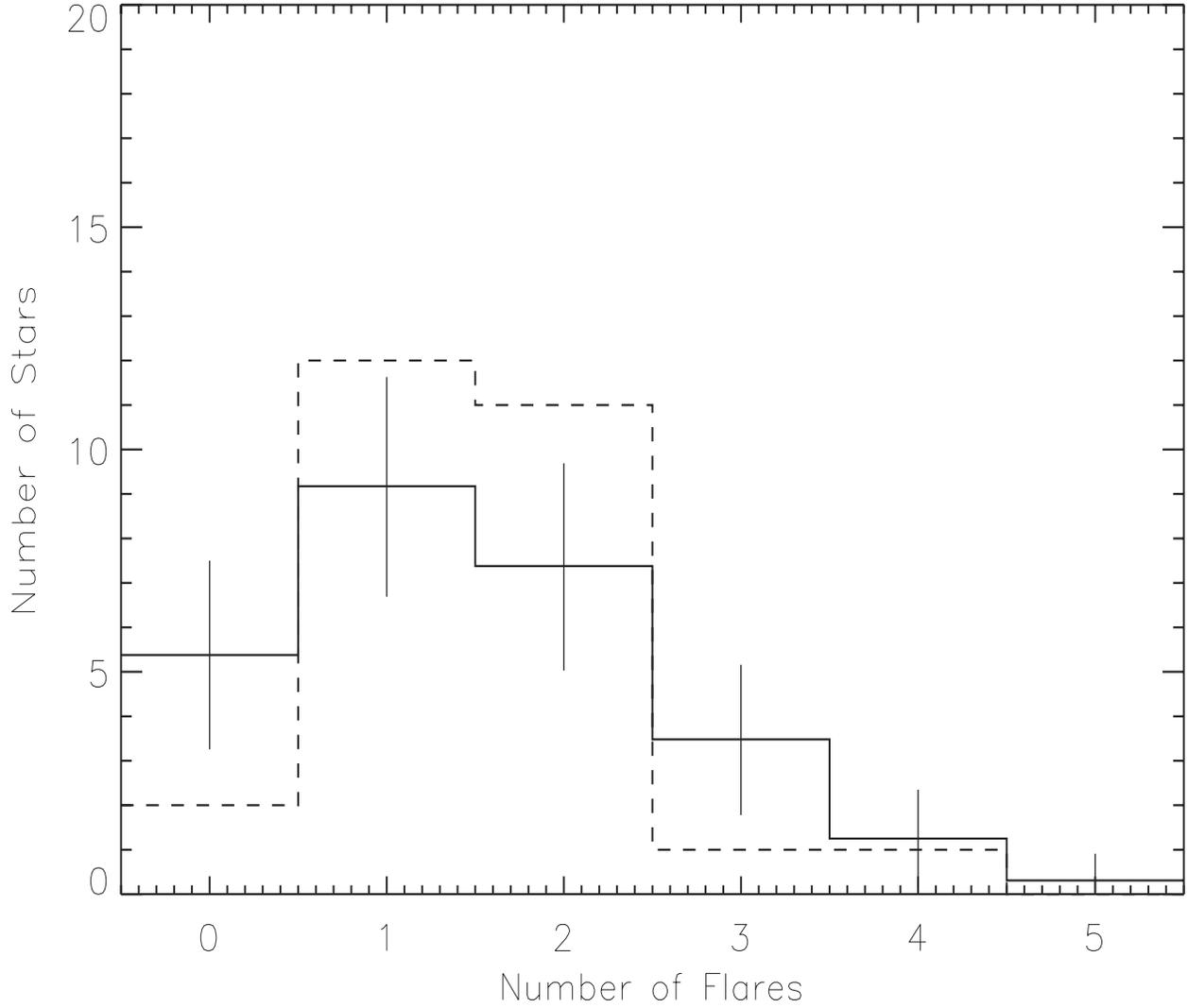}{0.0in}{90.}{400.}{470}{0.}{00.}
\caption{Distribution of flare frequency.  The solid histogram shows the
expected flare distribution if flaring is a random process occurring
once per 635 ks. The vertical lines show the standard deviation
observed in 1000 simulations. The true distribution is shown by the
dashed histogram.  The difference between the two distributions is not
statistically compelling.
\label{monte}}
\end{figure}  

\begin{figure}[tbph]
\plotfiddle{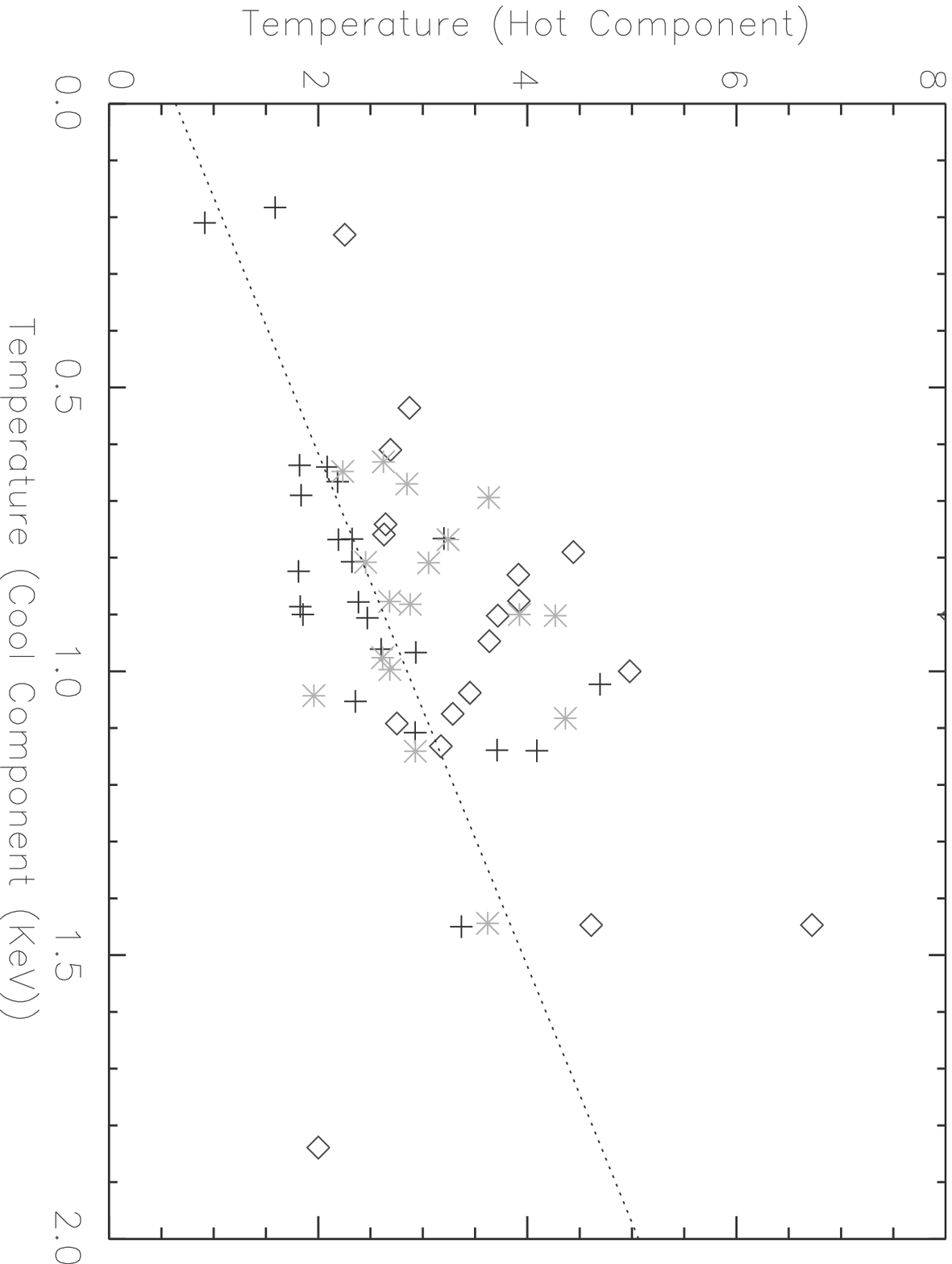}{0.0in}{90.}{400.}{470.}{0.}{-1.0in}
\caption{Scatter plot of the temperature of the hot coronal
component versus the cool component.
+= characteristic, *= elevated, diamond = average of each flare.
The dotted line is a fit to the characteristic data which are well
correlated.  Coronal components at
elevated and flare levels are not correlated. 
\label{kt1kt2}}
\end{figure}
\clearpage


\begin{deluxetable}{rccccccccccccc}
\tabletypesize{\tiny}
\tablecaption{Near-infrared and Optical Properties of Solar Mass Orion Nebula Cluster Stars  \label{observed}}
\tablewidth{0pt}
\tablehead{
\colhead{COUP} &\colhead{COUP J} &\colhead{JW$^a$}   & \multicolumn{2}{c}{offset(\arcsec)}   &
\colhead{V} &  \colhead{I}  & \colhead{J} &
\colhead{H}     & \colhead{K$_s$}  & 
\colhead{L} &\colhead{Spec.Type} &\colhead{A$_V$} \\
\colhead{} &\colhead{}  &\colhead{}  &\colhead{Optical} &\colhead{2MASS}  &\colhead{} }
\startdata
\object{17}  & 053443.0-052007 &63   &1.37&1.13&14.81&12.75 &11.17&10.37&10.09&---    &K6&1.58\\
\object{54}  & 053450.4-052020 &113  &0.03&0.43&16.43&14.20  &11.96&11.03&10.44&---    &K6&2.02\\
\object{57}  & 053450.7-052401 &116  &0.09&0.34&13.56&12.12 &11.01&10.53&10.28&---    &K5&0.33\\
\object{131} & 053458.8-052117 &187  &0.29&0.19&17.12&14.27 &11.98&10.95&10.24&---    &K5&3.95\\
\object{147} & 053500.4-052514 &198  &0.15&0.21&15.43&13.80  &12.19&11.19&10.40 &---    &K6&0.48\\
\object{177} & 053502.4-052046 &223a &0.08&0.19&16.06&13.64 &11.54&10.52&10.10 &---    &K5&2.85\\
\object{223} & 053504.7-051742 &253  &0.32&0.23&17.35&14.22 &11.53&10.10 &9.34 &---    &K5&4.66\\
\object{241} & 053505.4-052717 &268  &0.14&0.12&14.49&12.88 &11.8&11.08 &10.84&---    &K5-K6&0.43\\
\object{250} & 053505.7-052418 &278  &0.22&0.22&15.6 &13.66 &11.62&10.30 &9.33 &8.19   &K2-K7&1.61\\
\object{262} & 053506.2-052202 &286  &0.12&0.18&17.69&14.91 &11.66&10.07&9.30  &8.59   &K5&3.77\\
\object{314} & 053508.4-052829 &320  &0.18&0.24&17.1&14.73  &13.27&11.78&10.82&---    &K2&3.52\\
\object{515} & 053513.0-052030 &394  &0.20&0.23&18.82&14.98  &12.29&10.96&10.43&10.09  &K6&6.15\\
\object{567} & 053513.6-053057 &421  &0.07&0.26&12.94&11.48 &10.18&9.26 &8.62 &---    &K5$^{b}$&0.38\\
\object{753} & 053515.9-051459 &487  &0.16&0.21&14.57&12.79 &11.63&10.77&10.32&---    &K6&0.87\\
\object{828} & 053516.7-052404 &526b&0.35&0.16&13.77&11.87 &10.01&9.18 &8.89&8.84    &K2-K6&1.17\\
\object{1023}& 053519.2-052250 &9250&0.64&0.19&17.03&14.23 &11.98&10.86&10.36&9.88   &K5&3.82\\
\object{1127}& 053521.0-051637 &664 &0.14&0.23&16.93&14.05 &12.08&11.03&10.64&---    &K5.5-7&3.69\\
\object{1134}& 053521.0-053121 &673 &0.17&0.30&14.9 &13.07 &11.57&10.62&10.03&---    &K5&1.33\\
\object{1151}& 053521.3-052644 &683 &0.10&0.21&13.61 &11.76 &10.48&9.64 &9.40  &---    &K6&1.05\\
\object{1167}& 053521.7-052339 &694 &0.29&0.23&17.73&14.74 &12.54&11.42&10.80 &9.83   &K5-K7&3.97\\
\object{1235}& 053522.9-052241 &726 &0.23&0.20&19.24&15.35 &12.5 &11.03&10.33&9.95   &K5-K7&6.28\\
\object{1259}& 053523.6-052331 &738 &0.15&0.18&15.7 &13.84 &11.97&10.92&10.45&10.09  &K5&1.41\\
\object{1281}& 053524.2-052518 &750 &0.28&0.03&14.85&12.87 &11.53&10.54&9.94 &9.25   &K0-K5&1.72\\
\object{1326}& 053525.4-052134 &777 &0.64&0.02&18.46&15.15 &12.72&11.40 &10.58&---    &K6&4.79\\
\object{1327}& 053525.4-052135 &777 &0.59&0.02&18.46&15.15 &12   &10.90 &10.53&9.86   &K6&4.79\\
\object{1500}& 053532.9-051605 &892 &0.18&0.30&16.21&13.99 &11.62&10.49&10.06&---    &lateK&2.00\\
\object{1539}& 053537.5-052716 &930 &0.03&0.06&16.64&14.08 &12.71&11.94&11.71&---    &K5-K7&2.87\\
\object{1570}& 053542.4-052733 &962 &0.08&0.05&15.23&13.26 &11.39&10.57&10.12&---    &K6&1.35\\
\enddata
\tablenotetext{a}{Jones, B.~F.~\& Walker, M.~F.\ 1988, \aj, 95, 1755}
\tablenotetext{b}{Hillenbrand (1997) list this as a K5 (as measured by
Hillenbrand) with a previous spectral type of F8-G0III-IV with an
unknown reference}
\end{deluxetable}

\begin{deluxetable}{rcccccccr}
\tabletypesize{\scriptsize}
\tablecaption{Inferred properties of Solar Mass Stars  \label{tbl-2}}
\tablewidth{0pt}
\tablehead{
\colhead{COUP} &\colhead{log T$_{eff}$}  &\colhead{log L$_{bol}$}   & \colhead{Radius}   &
\colhead{Mass} &  \colhead{log Age}  & \colhead{$\Delta (I-K)$} &
\colhead{EW Ca}     &\colhead{Notes} \\
\colhead{} &\colhead{(K)}  &\colhead{(ergs s$^{-1}$)}   & \colhead{R$_{\odot}$}   &
\colhead{M$_{\odot}$} &  \colhead{(Yrs)}  & \colhead{mag.} &
\colhead{\AA}     &\colhead{}  }
\startdata
17&3.62&0.26&2.57&0.90&6.05&0.20&1.9&d\\
54&3.62&-0.21&1.49&0.92&6.73&1.20&-1.0&b\\
57&3.64&0.20&2.17&1.19&6.36&0.09&1.6&d\\
131&3.64&0.22&2.23&1.20&6.34&0.53&1.4&b\\
147&3.62&-0.43&1.16&0.90&7.09&1.96&-1.0&b,c\\
177&3.64&0.20&2.18&1.19&6.36&0.48&2.0&b\\
223&3.64&0.41&2.79&1.19&6.08&1.04&1.7&b\\
241&3.62&-0.07&1.75&0.93&6.50&0.16&2.8&d\\
250&3.64&-0.11&1.53&1.12&6.82&1.69&-9.6&a\\
262&3.64&-0.08&1.58&1.13&6.78&2.24&2.3&b\\
314&3.70&-0.08&1.25&1.10&7.28&0.98&0.3&b\\
515&3.62&0.49&3.32&0.90&5.79&-0.20&0.0&d\\
567&3.64&0.47&2.96&1.20&6.02&0.80&-3.5&a\\
753&3.62&0.07&2.06&0.91&6.29&0.28&1.8&d\\
828&3.62&0.52&3.43&0.90&5.76&0.74&1.2&b,e\\
1023&3.64&0.20&2.19&1.19&6.36&0.13&0.6&d\\
1127&3.62&0.26&2.55&0.90&6.05&-0.18&1.4&d,f\\
1134&3.64&0.06&1.86&1.17&6.57&0.81&1.3&b\\
1151&3.62&0.53&3.48&0.91&5.76&0.28&1.9&d\\
1167&3.62&0.05&2.01&0.91&6.32&0.18&-3.8&a,c\\
1235&3.62&0.37&2.90&0.90&5.92&0.17&0.0&d\\
1259&3.64&-0.23&1.33&1.05&7.00&&-0.3&\\
1281&3.64&0.24&2.27&1.20&6.31&0.38&0.0&b,c\\
1326&3.62&0.09&2.09&0.91&6.27&1.03&2.5&b\\
1327&3.62&0.09&2.09&0.91&6.27&1.03&2.5&b\\
1500&3.62&-0.13&1.63&0.94&6.60&&1.0&\\
1539&3.62&0.04&2.00&0.92&6.34&-0.71&5.2&d\\
1570&3.62&0.00&1.90&0.92&6.39&0.80&1.8&b\\
\enddata
\tablenotetext{a}{Accreting  disk}
\tablenotetext{b}{Disk - no evidence of accretion}
\tablenotetext{c}{Proplyd (Hillenbrand 1997) }   
\tablenotetext{d}{No evidence of a disk}
\tablenotetext{e}{JW 526a \& 526b}
\tablenotetext{f}{JW 892 \&3018}
\end{deluxetable}

\begin{deluxetable}{rrrccccc}
\tabletypesize{\normalsize}
\tablecaption{Time-averaged X-ray Luminosity of Solar Mass Stars \label{tbl-0}}
\tablewidth{0pt}
\tablehead{
\colhead{COUP}  &\colhead{NetCnts}   & \colhead{$\log L_s^a$}   &
\colhead{$\log L_h^a$}& \colhead{$\log L_t^a$}
& \colhead{$\log L_{t,c}^a$}  &\colhead{Notes}}
\startdata
  17   &    1083&  29.67&   29.72 & 29.99 & 30.07&c,g\\
  54   &    1583&  29.36&   29.26 & 29.61 & 29.86&\\
  57   &    4093&  30.28&   29.89 & 30.43 & 30.48&b\\
 131   &    9038&  29.96&   30.34 & 30.49 & 30.77&b,h\\
 147   &    2348&  29.51&   29.41 & 29.76 & 29.90&b,g\\
 177   &    5054&  29.79&   29.82 & 30.11 & 30.41&h\\
 223   &   10243&  30.00&   30.51 & 30.62 & 31.08&h\\
 241   &     314&  27.85&   29.24 & 29.26 & 29.94&\\
 250   &     497&  28.57&   29.16 & 29.26 & 29.45&\\
 262   &   11540&  30.06&   30.61 & 30.72 & 31.15&h\\
 314   &     458&  28.27&   29.34 & 29.38 & 29.96&\\
 515   &    4393&  29.66&   29.94 & 30.12 & 30.57&g,h\\
 567   &   10847&  30.18&   30.16 & 30.47 & 30.59&b\\
 753   &    5729&  29.87&   29.98 & 30.23 & 30.43&\\
 828   &   12067&  30.57&   30.89 & 31.06 & 31.20&d,f\\
1023   &    4852&  29.70&   29.95 & 30.14 & 30.53&f\\
1127   &5598&29.84  &  29.95     &  30.20  & 30.53 &g\\
1134   &4922&29.92  &  29.74     &  30.14  & 30.21 &g\\
1151   &   24094&  30.53&   30.41 & 30.77 & 30.85&g,h\\
1167   &     335&  28.30&   29.16 & 29.22 & 29.90&f\\
1235   &     345&  28.92&   29.27 & 29.43 & 29.71&b,d\\
1259   &    2183&  30.73&   30.79 & 31.06 & 31.26&e\\
1281   &    1554&  29.31&   29.25 & 29.58 & 29.72&f\\
1326   &     714&  29.18&   29.38 & 29.60 & 29.84&b,h\\
1327   &    1003&  29.35&   29.55 & 29.76 & 30.04&b,h\\
1500   &    4438&  30.20&   30.50 & 30.68 & 30.95&b,c,g\\
1539   &    1210&  29.33&   28.75 & 29.43 & 29.43&\\
1570   &   4146&  29.70&   29.91 & 30.12&  30.33&g \\
\enddata
\tablenotetext{a}{The logarithm of the luminosity in units of ergs s$^{-1}$.}
\tablenotetext{b}{Double sources (percentage of PSF extracted to avoid the
        neighbor): 57 (32\%), 131 (87\%), 147 (87\%), 567 (88\%),
        1259 (N/A), 1326 (44\%), 1327 (43\%), 1500 (67\%).  1326+1327
        are a pair of solar analogs (as the text notes) while the
        other companions have different or unknown masses.}
\tablenotetext{c}{PSF extends over detector edge,  the effect is minor       
and can be ignored.}
\tablenotetext{d}{Near a chip gap, the effect is significant for 828.}
\tablenotetext{e}{Suffers pileup only outer 7\% of photons are analyzed.}
\tablenotetext{f}{Suffers background contamination (usually PSF wings
of $\theta^{1}$\,Ori\,C).  This is unimportant as our sources are very strong.}
\tablenotetext{g}{Exhibits Ne/Fe spectral line abundance anomalies around
        0.8-1.3.}
\tablenotetext{h}{Spectral lines in the 1.3 keV region.}
\end{deluxetable}

\begin{deluxetable}{rrcccccccccc}
\tabletypesize{\tiny}
\tablecaption{Spectral Fits to Characteristic X-ray Flux \label{char_table}}
\tablewidth{0pt}
\tablehead{ \colhead{COUP} &\colhead{Duration} 
&\colhead{nbins}  & \colhead{$\chi^2/df$} &  \colhead{$\log P$}  
& \colhead{\nh}  & \colhead{KT1} & \colhead{KT2} &\colhead{E.M. Ratio} 
& \colhead{$\log L_{t,c}$}& \colhead{$\log E$} &\colhead{Notes}\\
\colhead{}&\colhead{(sec)} 
&\colhead{}  & \colhead{}  &  \colhead{}  
& \colhead{($\times 10^{21}$ cm$^{-2}$)}  & \colhead{(keV)} & \colhead{(keV)}& \colhead{(EM1/EM2)} 
& \colhead{(ergs s$^{-1}$)}& \colhead{(ergs)} &\colhead{}}
\startdata
17&549316&28&1.9&-2.4   &0.40&1.02&4.69&     1.57&     29.79&35.53&\\
54&650772&49&1.0&-0.4   &0.51&0.67&2.19&     1.14&     29.59&35.40&\\
57&735173&99&1.3&-1.5   &0.08&0.69&1.84&     0.72&     30.28&36.15&\\
131&396531&93&1.0&-0.4  &1.01&0.88&2.38&    0.31&     30.23&35.83&\\
147&523161&48&1.2&-0.8  &0.13&0.77&3.20&    0.20&     29.54&35.26&\\
177&789112&138&1.2&-1.0 &0.74&0.77&2.19&   0.60&     30.25&36.15&\\
223&578569&153&1.3&-1.9 &1.20&1.05&2.36&   0.19&     30.50&36.26&\\
241&744731&18&1.5&-0.9  &7.57&0.08&2.08&    $>$100&     33.23&39.10&a\\
250&556094&18&0.9&-0.2  &3.79&0.08&2.40&    $>$100&     32.41&38.16&a\\
262&519007&168&1.0&-0.5 &3.46&0.18&1.59&   54.83&    32.28&38.00&a\\
314&586693&16&1.4&-0.8  &5.94&0.11&2.33&    $>$100&     31.96&37.73&a\\
515&635948&118&1.3&-1.7 &1.57&0.77&2.32&   1.21&     30.38&36.19&\\
567&625851&153&1.5&---  &0.10&0.97&2.93&    0.44&     30.24&36.04&\\
753&640962&129&2.0&---  &0.22&1.11&2.93&    0.27&     30.06&35.87&\\
828&632461&187&1.8&---  &0.30&1.14&4.09&    0.17&     30.85&36.65&\\
1023&830706&163&1.2&-1.3&0.98&0.91&2.47&  0.23&     30.29&36.21&\\
1127&734165&133&1.1&-0.6&0.94&0.82&1.81&  0.52&     30.30&36.16&\\
1134&669494&109&1.4&-2.5&0.34&0.64&2.09&  0.75&     30.06&35.88&\\
1151&689861&232&1.8&--- &0.28&0.81&2.32&  0.42&     30.72&36.56&\\
1167&527616&9&1.2&-0.5  &1.02&$> 15$&$>15$& 0.10&     28.96&34.68&\\
1235&817070&19&1.6&-1.1 &1.62&1.50&$>15$&  15.15&    29.66&35.58&\\
1259&572943&44&0.9&-0.2 &0.37&0.89&1.83&   0.36&     31.85&37.60&\\
1281&797023&68&1.1&-0.5 &0.45&0.96&2.60&   0.76&     29.60&35.50&\\
1326&746680&30&1.1&-0.5 &1.23&0.64&1.82&   0.33&     29.79&35.66&\\
1327&825748&48&1.0&-0.3 &1.23&0.90&1.85&   0.39&     29.95&35.87&\\
1500&618698&109&1.9&--- &0.51&1.45&3.37&   0.05&     30.54&36.33&\\
1539&757046&38&1.2&-0.6 &0.74&0.21&0.92&   10.09&    29.95&35.83&\\
1570&561651&75&1.6&-3.0 &0.35&1.14&3.71&   0.27&     29.81&35.56&\\
\enddata
\tablenotetext{a}{Hydrogen column \nh was fit as a free parameter.}
\end{deluxetable}

\begin{deluxetable}{rrcccccccccc}
\tabletypesize{\tiny}
\tablecaption{Spectral Fits to Elevated X-ray Flux \label{tbl_e1}}
\tablewidth{0pt}
\tablehead{\colhead{COUP}&\colhead{Duration} 
&\colhead{nbins}  & \colhead{$\chi^2/df$} &  \colhead{$\log P$}  
& \colhead{\nh}  & \colhead{KT1} & \colhead{KT2} &\colhead{E.M. Ratio} 
& \colhead{$\log L_{t,c}$}& \colhead{$\log E$} &\colhead{Notes}\\
\colhead{}&\colhead{} &\colhead{(sec)} &  \colhead{}   &\colhead{}
& \colhead{($\times 10^{21}$ cm$^{-2}$)}  & \colhead{(keV)} & \colhead{(keV)}& \colhead{(EM1/EM2)} 
& \colhead{(ergs s$^{-1}$)}& \colhead{(ergs)} &\colhead{}}
\startdata
17&204741&26&1.06&-0.4   &0.40&0.90&3.92&    0.50&  30.22&35.54&\\
54&35090&15&1.54&-1.0    &0.51&0.65&2.23&     0.31&  29.83&34.38&\\
57&42944&34&0.82&-0.1    &0.08&0.81&3.06&     3.06&  30.60&35.23&\\
131&99546&106&0.93&-0.2  &1.01&0.88&2.68&   0.63&  30.53&35.53&\\
147&231582&50&0.93&-0.2  &0.13&0.67&2.85&   0.11&  29.85&35.22&\\
177&52790&33&1.06&-0.4   &0.74&0.90&4.27&    0.45&  30.51&35.24&\\
223&32474&116&1.21&-1.2  &1.20&1.14&2.93&   0.30&  30.73&35.24&\\
262&76306&187&1.03&-0.4  &2.33&0.98&2.62&   0.48&  31.15&35.03&a\\
515&131208&68&1.28&-1.2  &1.57&0.81&2.45&   0.27&  30.61&35.72&\\
567&22858&95&1.30&-1.5   &0.10&1.08&4.36&    2.15&  30.51&34.87&\\
753&61711&91&1.55&-3.0   &0.22&1.44&3.62&    0.12&  30.33&35.13&\\
828&158275&176&1.13&-0.9 &0.30&12.09&3.71& 0.45&  31.23&36.43&\\
1023&18806&17&0.49&0.0   &0.98&0.69&3.63&    0.04&  30.59&34.87&\\
1127&22663&43&1.08&-0.5  &0.94&1.04&1.96&   2.66&  30.50&34.86&\\
1134&75494&49&1.44&-1.6  &0.34&0.63&2.62&   0.80&  30.33&35.21&\\
1151&132262&169&1.16&-1.1&0.27&0.88&2.88& 0.00&  30.94&36.06\\
1259&65114&13&0.66&-0.1  &0.37&2.23&2.29&   0.32&  30.99&35.80\\
1281&52489&13&0.65&-0.1  &0.45&0.77&3.24&   0.67&  30.01&34.73\\
1326&72598&12&1.24&-0.6  &1.23&1.24&78.41&  0.45&  30.09&34.95\\
1500&15234&64&1.83&---   &0.51&0.08&6.14&    0.23&  30.90&35.08\\
1570&65831&31&1.16&-0.6  &0.35&1.00&2.69&   0.00&  30.13&34.95\\
\enddata
\tablenotetext{a}{Hydrogen column \nh was fit as a free parameter.}
\end{deluxetable}

\begin{deluxetable}{rrrrccccccccc}
\tabletypesize{\tiny}
\tablecaption{Spectral Fits to Integrated Flares \label{tbl-flares}}
\tablewidth{0pt}
\tablehead{ \colhead{COUP}& \colhead{Flare \#} &\colhead{Duration} 
&\colhead{nbins}  & \colhead{$\chi^2/df$} &  \colhead{$\log P$}  
& \colhead{\nh}  & \colhead{KT1} & \colhead{KT2} &\colhead{E.M. Ratio}
& \colhead{$\log L_{t,c}$}& \colhead{$\log E$} &\colhead{Notes}\\ 
\colhead{}&\colhead{} &\colhead{(sec)}  &\colhead{}  & \colhead{}   &\colhead{}
& \colhead{($\times 10^{21}$ cm$^{-2}$)}  & \colhead{(keV)} & \colhead{(keV)}& \colhead{(EM1/EM2)} 
& \colhead{(ergs s$^{-1}$)}& \colhead{(ergs)} &\colhead{}}
\startdata
17&1&44113&13&1.26   &-0.6&0.40&0.90&4.37& 0.19&        30.46&35.11&a\\
54&1&9067&10&1.90    &-1.1&0.51&0.08&2.08& 0.35&       30.55&34.51&a\\
54&2&154578&35&0.62  &0.0 &0.51&0.61&2.69& $>$100&       30.07&35.26&b\\
57&1&58063&34&1.24   &-0.8&0.08&0.88&3.92& 0.30&       30.77&35.53&a\\
57&2&13326&15&0.76   &-0.2&0.08&0.72&$>15$&0.73&       30.99&35.12&b\\
131&1&190371&134&0.95&-0.2&1.01&0.18&3.06& $>$100&       30.81&36.09&b\\
131&2&163058&154&0.90&-0.1&1.01&0.90&3.72& 0.11&       30.92&36.13&b\\
147&1&24873&11&0.92  &-0.3&0.13&0.76&2.55& 0.45&       29.90&34.30&g\\
147&2&7864&11&1.24   &-0.6&0.13&1.14&$>15$&0.28&       30.25&34.15&c,g\\
177&1&7604&15&0.79   &-0.2&0.74&0.90&8.06& 0.35&       30.86&34.74&a\\
223&1&173205&173&0.99&-0.3&1.20&0.08&4.77& 0.00&       31.02&36.25&a\\
223&2&65264&96&1.40  &-2.2&1.20&0.08&9.56& 0.37&       31.11&35.92&a\\
241&1&104787&14&0.56 &-0.1&11.26&0.08&1.59& $>$100&       35.18&40.20&b,g\\
250&1&246439&16&0.79 &-0.2&2.96&0.09&2.10& $>$100&       32.15&37.54&a,g\\
262&3&254193&214&0.95&-0.2&2.29&1.13&3.17& 3.28&       31.23&36.63&a,g\\
314&1&99223&13&1.26  &-0.6&4.94&0.13&2.61& $>$100&       31.87&36.86&a,g\\
314&2&14226&&&&&&&                         &                & &c,e\\
314&3&131465&11&1.03 &-0.4&20.34&0.08&0.58&$>$100&       37.56&42.68&b,c,f\\
314&4&17910&&&&&&&                             &                & &a,e\\
515&1&82357&54&0.90  &-0.2&1.57&0.76&2.63& 0.38&       30.79&35.70&a\\
567&1&34980&50&1.14  &-0.6&0.83&0.79&4.44& 0.41&       30.96&35.50&a\\
567&2&67095&66&1.25  &-1.0&0.10&1.00&4.98& 0.19&       30.60&35.43&a,i\\
567&3&98734&112&0.96 &-0.2&0.20&1.04&3.45& 0.41&       30.74&35.73&b\\
753&1&80367&43&1.55  &-1.8&0.22&1.45&4.61& 0.21&       30.36&35.26&a\\
753&2&66472&58&1.12  &-0.6&1.63&0.23&2.25& 6.31&       31.44&36.26&a\\
1127&1&52594&32&1.28 &-0.8&0.94&1.09&2.75& 0.10&       30.57&35.29&a,h,i\\
1127&2&40090&60&0.69 & 0.0&0.94&0.54&2.87& 0.55&       31.03&35.63&a\\
1134&1&19758&16&0.87 &-0.2&0.37&0.83&3.92& 0.96&       30.57&34.86&a\\
1134&2&35233&24&0.67 &-0.1&0.37&0.74&2.65& 0.31&       30.48&35.03&a\\
1151&1&27396&69&1.19 &-0.9&0.27&0.95&3.64& 0.36&       31.03&35.47&b\\
1167&1&203281&15&0.74&-0.2&5.33&0.72&5.95& 5.47&       30.38&35.69&a\\
1235&1&32443&&&&&&&                            &              & &a,e\\
1259&1&149468&25&1.37&-0.9&0.37&1.84&2.00& 0.82&       32.40&37.57&b,i\\
1259&2&61987&62&1.15 &-0.7&0.37&1.45&6.72& 4.41&       33.19&37.98&a\\
1326&1&4103&&&&&&&                                        & & &c,e\\
1326&2&26132&13&1.22 &-0.6&1.23&0.08&2.32& $>$100&       31.05&35.47&a\\
1327&1&16381&10&1.98 &-1.2&1.23&0.08&4.46& 0.00&       30.72&34.94&a\\
1500&1&215580&101&1.78&-   &0.51&4.96&$>15$&$>$100&      30.97&36.30&a\\
1539&1&70264&27&1.23 &-0.7&0.74&0.20&1.68& 7.95&       30.76&35.60&a\\
1570&1&41441&16&0.54 &-0.1&0.35&1.08&3.29& 0.17&       30.21&34.82&a,e\\
1570&2&94762&64&1.00 &-0.3&0.35&4.92&$>15$&$>$100&       30.54&35.52&b\\
\enddata
\tablenotetext{a}{Flare morphology - Rapid Rise followed by exponential decay.}
\tablenotetext{b}{Flare morphology - Symmetric}
\tablenotetext{c}{Flare morphology - Spike, Rise and fall within 5ks.}
\tablenotetext{d}{Flare morphology - weak irregular.}
\tablenotetext{e}{Not enough photons to fit.}
\tablenotetext{f}{Formally good fit but very few photons.}
\tablenotetext{g}{Nearby star may have affected fit.}
\tablenotetext{h}{Duration overstated, probably closer to 5 ks.}
\tablenotetext{i}{Hydrogen column \nh was fit as a free parameter.}
\end{deluxetable}

\begin{deluxetable}{rrrrcccccccccc}
\tabletypesize{\tiny}
\tablecaption{Spectral Fits to the Peak of Each Flare \label{tbl-brightest}}
\tablewidth{41pc}
\tablehead{ \colhead{COUP}& \colhead{Flare \#} &\colhead{Duration} 
&\colhead{nbins}  & \colhead{$\chi^2/df$} &  \colhead{$\log P$}  
& \colhead{\nh}  & \colhead{KT1} & \colhead{KT2} &\colhead{E.M. Ratio} 
& \colhead{$\log L_{t,c}$}& \colhead{$\log E$}\\
\colhead{}&\colhead{} &\colhead{(sec)}  &\colhead{}  & \colhead{}   &\colhead{}
& \colhead{($\times 10^{21}$ cm$^{-2}$)}  & \colhead{(keV)} & \colhead{(keV)}& \colhead{(EM1/EM2)} 
& \colhead{(ergs s$^{-1}$)}& \colhead{(ergs)}}
\startdata
17&1&24133&18&0.82&-0.2  &0.40&0.08&2.92&23.57&30.62&35.01\\
54&1&9066&10&1.90&-1.1   &0.51&0.08&2.08&$>$100&30.55&34.51\\
54&2&85122&26&0.99&-0.3  &0.51&0.50&3.02&0.26&30.20&35.13\\
57&1&10979&16&0.37&0.0   &0.08&0.71&4.77&0.15&31.01&35.05\\
57&2&13326&15&0.76&-0.2  &0.08&0.72&$>$15&0.30&30.99&35.12\\
131&1&20013&49&0.99&-0.3 &1.01&0.11&4.88&9.85&31.43&35.74\\
131&2&24266&42&1.05&-0.4 &1.01&0.71&4.15&0.17&31.08&35.46\\
147&1&21775&15&0.63&-0.1 &0.13&0.74&$>$15&0.67&29.98&34.32\\
147&2&7864&11&1.24&-0.6  &0.13&1.14&$>$15&0.28&30.25&34.15\\
177&1&1848&14&0.43&0.0   &0.74&1.08&$>$15&0.17&31.20&34.47\\
223&1&9407&59&1.14&-0.7  &1.20&$>$15&$>$15&2.57&31.71&35.68\\
223&2&25817&60&2.06&---  &1.20&0.08&$>$15&0.00&31.28&35.69\\
250&1&7473&10&2.79&-2.0  &0.41&0.08&$>$15&0.00&30.30&34.18\\
262&3&965&17&1.38&-0.8   &0.96&0.08&$>$15&0.00&31.60&34.59\\
314&1&19279&14&3.25&---  &0.90&0.08&$>$15&0.00&30.02&34.31\\
515&1&18945&23&1.01&-0.3 &1.57&1.78&5.47&3.10&30.97&35.25\\
567&1&10862&27&2.54&---  &0.10&0.08&$>$15&0.00&31.06&35.10\\
567&2&12673&23&1.17&-0.6 &0.10&1.12&6.95&0.14&30.85&34.95\\
567&3&31332&67&0.85&-0.1 &0.10&1.14&3.59&0.34&30.94&35.44\\
753&1&4022&19&0.80&-0.2  &0.22&2.32&$>$15&0.58&30.82&34.43\\
753&2&4558&19&3.60&---   &0.22&0.08&$>$15&0.00&31.33&34.99\\
1127&1&2811&13&0.59&-0.1 &0.98&0.75&4.31&0.42&31.00&34.44\\
1127&2&8231&30&0.72&-0.1 &0.98&1.72&5.95&0.52&31.40&35.32\\
1134&1&6395&13&0.82&-0.2 &0.34&0.16&3.03&2.42&30.93&34.74\\
1134&2&15383&14&1.45&-0.8&0.34&0.71&4.10&0.39&30.60&34.79\\
1134&3&49539&40&0.66&0.0 &0.34&0.65&3.22&0.31&30.59&35.29\\
1151&1&3078&17&1.21&-0.6 &0.27&1.12&$>$15&0.35&31.40&34.89\\
1259&1&55286&14&0.36&0.0 &0.37&1.55&$>$15&40.62&31.26&36.01\\
1259&2&2926&16&2.94&---  &0.37&3.26&3.59&12.69&32.60&36.06\\
1327&1&16380&10&1.98&-1.2&1.23&0.08&4.46&0.00&30.72&34.94\\
1539&1&32052&16&0.33&0.0 &0.74&0.17&1.54&12.89&30.81&35.31\\
1570&1&8585&15&0.86&-0.2 &0.35&1.83&$>$15&0.60&30.43&34.36\\
1570&2&4836&17&1.01&-0.4 &0.35&9.87&$>$15&0.00&31.09&34.78\\
\enddata
\end{deluxetable}

\end{document}